\documentclass[a4paper,journal]{IEEEtran}

\usepackage{url}
\usepackage{amsmath}
\usepackage{amsthm}
\usepackage{amssymb}
\usepackage{cite}
\usepackage{algorithmic}
\usepackage{algorithm2e}
\usepackage[caption=false,font=footnotesize,farskip=7pt]{subfig}
\usepackage{graphicx}
\usepackage{float}
\usepackage{color}
\usepackage{tikz}
\usepackage{booktabs}
\usepackage{fancybox}
\usepackage{ulem}
\usepackage[utf8x]{inputenc}
\usepackage{lipsum}
\usepackage[breaklinks]{hyperref}

\def\R{{\mathbb R}}
\def\T{{\mathbf T}}
\def \M {\text{TiSANCR}}
\def \Q {\mathbf{Q}}
\def \K {{\mathbf K}}
\def \V {{\mathbf V}}
\def \W {{\mathbf W}}

\def \Uset {\mathcal{U}}
\def \Vset {\mathcal{V}}
\def \Tset {\mathcal{T}}
\def \uv {\boldsymbol u}
\def \vv {\boldsymbol v}
\def \tv {\boldsymbol t}
\def \bv {\boldsymbol b}
\def \lv {\boldsymbol{\ell}}

\begin{document}

\title{Time-aware Self-Attention Meets Logic Reasoning in Recommender Systems}

\author{\IEEEauthorblockN{Zhijian Luo, 
Zihan Huang, 
Jiahui Tang,
Yueen Hou and
Yanzeng Gao\\}
\IEEEauthorblockA{School of Computer, Jiaying University, \\
Meizhou, R. P. China, 514015\\}
}

\maketitle

\begin{abstract}
At the age of big data, recommender systems have shown remarkable success as a key means of information filtering in our daily life.
Recent years have witnessed the technical development of recommender systems, from perception learning to cognition reasoning which intuitively build the task of recommendation as the procedure of logical reasoning and have achieve significant improvement.
However, the logical statement in reasoning implicitly admits irrelevance of ordering, even does not consider time information which plays an important role in many recommendation tasks.
Furthermore, recommendation model incorporated with temporal context would tend to be self-attentive, i.e., automatically focus more (less) on the relevance (irrelevance), respectively.

To address these issues, in this paper, we propose a Time-aware Self-Attention with Neural Collaborative Reasoning (\M) based recommendation model, which integrates temporal patterns and self-attention mechanism into reasoning-based recommendation.
Specially, temporal patterns represented by relative time, provide context and auxiliary information to characterize the user's preference in recommendation, while self-attention is leveraged to distill informative patterns and suppress irrelevances.
Therefore, the fusion of self-attentive temporal information provides deeper representation of user's preference.
Extensive experiments on benchmark datasets demonstrate that the proposed \M~achieves significant improvement and consistently outperforms the state-of-the-art recommendation methods.
\end{abstract}


%

\section{Introduction}

Over the past of decades, the explosive growth of data has produced large number of redundant information, leading to information overload \cite{gantz2012digital} and making it difficult for individuals to acquire information that meet their satisfaction.
Recommender system (RS) has become prevalent technique of information filtering for alleviating this issue, and has achieved tremendous success in many fields, such as social network, e-learning, e-tourism, e-commerce, etc\cite{lu2015recommender}.
RS learns user's previous interactions with items (e.g., ratings, clicks, add-to-favorite, dislike and purchase) and automatically predicts user's preferences to recommend the item of interest \cite{bobadilla2013recommender}.
%

Collaborative filtering (CF) is a dominant methodology
which relies on the information provided by users who share same preference, i.e., leveraging the wisdom of crowd.
In particular, embedding based models, such as the latent factor model (LFM) \cite{lu2012recommender},
constructs the low-rank embeddings,
and predicts the utility of interaction usually formulated as a linear kernel, i.e., a dot product of the embeddings.
However, trivially combining the multiplication of latent embeddings linearly, the inner product may not be sufficient to capture the complex structure of user-item interactions well.

Theoretically, traditional matrix factorization is identical to a two-layer neural network, in which the latent layer learns low-rank embeddings of users and items, and inner product operation are conducted in the second layer.
Recently, deep neural networks (DNN) have been further extended to CF methods, which deploy complex structures of neural networks to improve performance of RS, including AutoRec\cite{sedhain2015autorec}, Neural Collaborative Filtering \cite{he2017neural}, CDAE\cite{wu2016collaborative}, CDL \cite{wang2015collaborative}, DeepCoNN \cite{zheng2017joint}, to name a few.
However, it is controversial whether complex architecture of DNN is superior to simple matching function \cite{resnick1994grouplens, dacrema2019we, dacrema2021troubling,ferrari2020critically}. %

Though most of existing works have obtained impressive performance in recommendation, they are still designed for leaning correlative patterns for data and making prediction based on the idea of matching.
A recent line of researches including NCR\cite{chen2021neural} and LINN\cite{shi2020neural}, focus on reasoning-based recommendation models, in which the procedure of recommendation exhibit certain expressions of propositional logic.
In this scenario, cognitive reasoning is capable of modeling complex relationships between users and items in terms of propositional logic.

Reasoning-based recommendations achieve significant improvement on the ranking performance\cite{chen2021neural,shi2020neural}, however, the expressions of propositional logic modeled by neural modules have not taken time information of interactions into consideration.
On one hand, the logic expressions including conjunction and disjunction operations, neglect the ordering of events which is prominent in many recommendation applications \cite{li2020time,kang2018self,ying2018sequential}.
For example, the conjunction $a \wedge b$ which represents that one user bought item $a$ and $b$, implicitly admits irrelevance of ordering due to the commutative law of conjunction, i.e., $a \wedge b \Leftrightarrow b \wedge a$.
However, the ordering of purchase list should have great impacts on the results of recommendation.
For instance, if a user purchased T-shirt and then bought heels after three months, one would expect the user to purchase socks to make collocation to heels.
In contrast, leisure shorts would be recommended to the user if he/she purchased heels and bought T-shirt three months later.
Temporal recommendation \cite{koren2009collaborative, kang2018self, wu2017recurrent, xiong2010temporal} alleviates these issues by exploiting temporal information for recommendation of candidate items, however, researches on the integration of cognitive logic and temporal information in recommender system have received little to no attention.

\begin{figure*} [!t]
  \centering
  \includegraphics[width=0.9\linewidth]{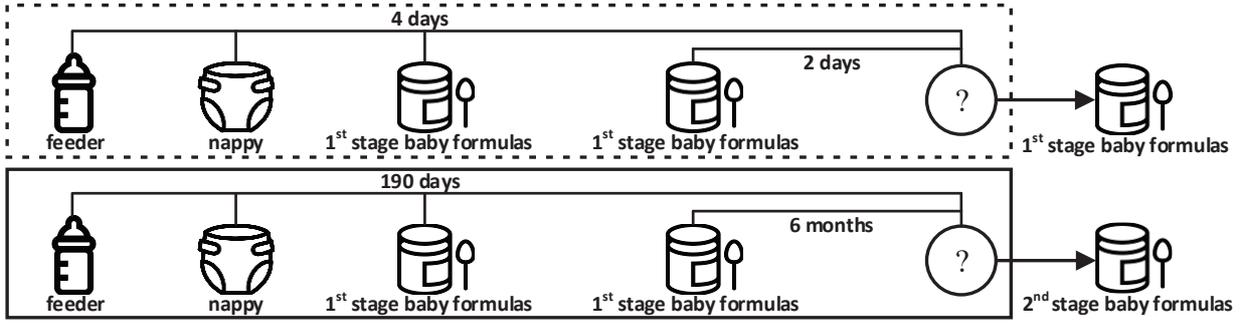}
  \caption{An example of self-attention on temporal information in recommendation system for a mother of newborn.}\label{fig:running-example}
\end{figure*}

On the other hand, temporal recommendation\cite{koren2009collaborative,ye2020time} typically model the temporal dynamics of user's preference, which characterizes user time-sensitive behaviors.
However, exploiting temporal information in recommendation, to some extern, does not provides highlighted correlations of interactions in context.
As exampled in Fig. \ref{fig:running-example}, the purchase list of a mother of newborn is characterized in pairs as (item, time) in which the term of time is defined as relative time.
In the upper dashed rectangle,
the historical interactions are (feeder, 4 days), (nappy, 4 days), (1st stage baby formulas, 4 days) and (1st stage baby formulas, 2 days),
one should expect that the mother purchase 1st stage baby formulas with high probability.
On the contrary, as shown in the lower rectangle,
the historical interactions are (feeder, 190 days), (nappy, 190 days), (1st stage baby formulas, 190 days) and (1st stage baby formulas, 6 months),
2nd stage baby formulas should be more reasonable to recommend since that 2nd stage baby formulas is suitable for 6 month old baby.
Therefore in this example, the relative time 6 months provides significant information in context to the recommendation of 2nd stage baby formulas.
Under this condition, the model of RS would tend to be self-attentive on temporal information.

In order to enhance the influence of temporal information on recommendation, the recent advance in DNN model - the attention mechanism, is commonly employed in recommendation system \cite{li2020time,kang2018self,shaw2018self,chen2017attentive,xiao2017attentional}.
The self-attention mechanism provides guidance to model to distill informative patterns and suppress irrelevance.
Intuitively, the task of recommendation based on previous interactions is essentially similar to the mechanism of self-attention, where the output of model depends on relevant parts of the input.
In reasoning-based recommendation, the output of logical reasoning, e.g., the logical value of material implication $(a\wedge b) \to c $ which indicates whether the item $c$ is recommended or not when the interactions $a$ and $b$ were observed, could be heavily dependent on the parts of observations $a$ and/or $b$.
Hence, it is an open question \textit{whether integrating temporal information and self-attention into reasoning-based recommendation system contributes to improving performance?}

In this paper, we answer this question to fulfill this research gap by proposing \M ~(Time-aware Self-Attention with Neural Collaborative Reasoning) based recommendation system, which bridges self-attentive temporal information to the NCR framework.
Specifically, in order to capture the temporal dynamics of logic reasoning in recommendation, we extract the relative time of each interactions as context information, which highlights how temporal dynamics have impact on the relationship between historical interactions and candidate  item.
Notably, the logical reasoning in recommendation should depend on temporal information in the past of `relevant' interactions which the model should concentrate on successively (as exampled in Fig. \ref{fig:running-example}).
In order to enhance the influence of temporal information on recommendation, we build self-attention module to adaptively capture weights of previous interactions in terms of timestamp.
Furthermore, employing attention mechanism in reasoning-based recommendation is beneficial to interpret the logical reasoning in RS.
The main contributions of this work are summarized as follows.
\begin{enumerate}
  \item \noindent We propose a time-aware self-attention with NCR (\M) based recommendation system, which integrates temporal information of user-item interactions to reasoning-based recommendation and exploits self-attention to capture the impacts of temporal information on logical reasoning.
  \item To the best of our knowledge, this work is the first to introduce temporal information and self-attention mechanism to build reasoning-based recommendation model, which improve the performance and interpretation of the recommendation system.
  \item Experimental results on benchmark datasets show that our \M~model achieves significant improvement and consistently outperforms the state-of-the-art baselines.
\end{enumerate}

The remainder of this paper is organized as follows. In Section \ref{sec_related} we briefly revisit some researches relevant to our work in the literature, and the proposed \M ~model is presented in Section \ref{sec_proposed}.
In Section \ref{sec_exp}, we conduct experiments to compare our \M ~model with state-of-the-art baselines over public datasets.
At last, we provide the conclusion and future work of this study in Section \ref{sec_conclusion}.

\section{Related works} \label{sec_related}
In this section, we briefly review several lines of work that are mostly related to ours, which would further highlight the contributions of our method.
We first discuss temporal recommendation, followed by reasoning-based recommendation, and at last we discuss the attention mechanism in RS.

\subsection{Temporal Recommendation}
Typically in recommender systems, temporal modeling of interaction is important in many tasks such movie recommendation \cite{koren2009collaborative,xiong2010temporal}, news recommendation \cite{li2014modeling,liu2010personalized} and music recommendation\cite{koenigstein2011yahoo}, etc.
As an important contextual information, temporal information is usually considered to be associated with historical behaviors to adapt the recommendations accordingly\cite{campos2014time}.
Temporal model of recommender system has been proven to significantly improve the performance over non-temporal model, hence attracting much attention from both the industrial and academic research communities \cite{ye2020time,koren2009collaborative}.

Incorporation of temporal information has been already applied in CF-based methods to prove its effectiveness.
The representative work is TimeSVD++ \cite{koren2009collaborative}, where temporal dynamics is leveraged to extend the matrix factorization model.
The time-dependent factors are constrained to be similar to the one in previous time step, i.e., to maintain the smoothness over time.
A tensor factorization approach which extends matrix factorization was proposed in BPTF \cite{xiong2010temporal} to capture the temporal dynamics of user's preference, where temporal information is modeled as an additional dimension.
In order to predict next step from previous data, a temporal matrix factorization approach was introduced in \cite{yu2016temporal} where the matrix factorization was viewed as an auto-regressive model.
In \cite{yuan2013time}, temporal pattern similarity between users was incorporated to extend the similarity function in CF.

Recently, deep leaning approaches for recommendation system, especially recurrent neural network (RNN) have achieved greater improvement \cite{pei2017interacting, sun2018attentive,yu2016dynamic}.
In JODIE \cite{kumar2019predicting}, the RNNs of user and item were developed to update the embedding of user and item respectively.
MTAM \cite{li2020time}, TiSASRec \cite{ji2020sequential}, and CTA \cite{wu2020deja} considered to use time interval as temporal information.
To model user's long-term and short-term interest, a multi-rate temporal model was proposed in TDSSM \cite{song2016multi}.
The absolute and relative temporal patterns ware modeled jointly into attention models in TASER \cite{ye2020time}.
To model the evolution of user preference,  in TGSRec \cite{fan2021continuous}, a temporal collaborative transformer layer was proposed, while the sequential patterns and temporal collaborative signals are unified in graph collaborative transformer.
However, researches on the integration temporal information and cognitive reasoning  in recommender system have received little to no attention.

\subsection{Reasoning-based Recommendation}
Reasoning-based recommender systems including NCR\cite{chen2021neural} and LINN\cite{shi2020neural}, claim that recommendation should require not only the ability of pattern recognition, but also the ability of cognitive reasoning in terms of data.
The procedure of recommendation based on historical events of interactions, e.g., if the user purchased $r$ items $\{v_1,v_2,\cdots,v_r\}$ then $v_x$ would be recommended with high probability, exhibits certain expressions of propositional logic below
\begin{equation}
  \left( v_1 \wedge v_2 \wedge \cdots \wedge  v_r \right)  \to v_x.
\end{equation}
According to De Morgan's Law, the expression above can be rewritten into the following statement
\begin{equation}\label{eq:ncr_logic_example}
  \neg v_1 \vee \neg v_2 \vee \cdots \vee \neg v_r \vee v_x,
\end{equation}
which only contains two basic logical operations $\neg$ and $\vee$.
In this scenario, cognitive reasoning is capable of modeling complex relationships between users and items in term of propositional logic.

However, it is problematic for logic reasoning to search optimum of recommendation in symbolic space which is not differential.
What's more, hand-crafted rules of logic reasoning would degrade the generalization performance for the task of recommendation.
To address these issues, NCR\cite{chen2021neural} modeled the basic logic expressions (including negation $\neg$ and disjunction $\vee$) as neural modules, which integrate the powers of representation learning and logic learning.
In particular, representations capture the similarity patterns and logic provides cognitive reasoning for recommendation in a shared neural architecture, making it feasible for optimization and inference.
Furthermore, the neural architecture of logic reasoning is trained collaboratively in logical space, in which each user contributes their logical expression.
Our work follows this pipeline but contributes in that we incorporate temporal patterns of interaction between user and item into reasoning-based recommendation, which is considered to provide deeper representation of user preference in reasoning-based recommendation.

\begin{figure} [!t]
  \centering
  \includegraphics[width=\linewidth]{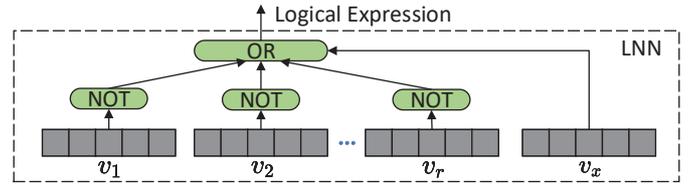}
  \caption{Logical Neural Network (LNN) of Eq. \eqref{eq:ncr_logic_example}}
  \label{fig:lnn-module}
\end{figure}

The corresponding logic neural network (LNN) module of Eq.\eqref{eq:ncr_logic_example} is manifested in Fig. \ref{fig:lnn-module}.
The inputs of LNN are the vectorization of each historical item \{$v_1$,$v_2$,$\cdots$,$v_r$\} and candidate item $v_x$, while the logical operations of negation and disjunction (i.e., NOT module and OR module in Fig. \ref{fig:lnn-module}) are implemented by two-layer MLPs respectively.
The output of LNN, i.e. logical expression represented as vector, is further compared to the constant true vector.

\subsection{Attention mechanism}
Attention mechanism has been shown effective in many machine learning tasks such as machine translation, image/video captioning and visual tracking, among others\cite{ wang2017residual ,zhai2006visual ,luong2015effective}. 
Essentially, the key idea behind attention mechanism is that the model focus more on the relevant elements of input, and less on the irrelevant elements which may suppress the impact of the relevant ones.
In particular, the importance evaluation of the elements are realized by assigning different correlation scores to different parts of input, and automatically highlighting the elements most related to the task so as to improve the performance.
Formally, the attention mechanism employs the scaled dot-product attention defined as follow
\begin{equation}\label{eq:self_att}
	\text{Attention}(\Q,\K,\V) = \text{softmax} \left( \frac{\Q^T \K}{\sqrt{d}} \right) \V
\end{equation}
where $\Q$, $\K$, $\V \in \R^{d \times d}$ are query, key and value respectively, $\Q^T$ is the transpose matrix of $\Q$.
The core component of attention mechanism is the self-attention module where $\Q$, $\K$ and $\V$ take identical object as input.

The developments of attention mechanism have been further extended to recommender system \cite{chen2017attentive,xiao2017attentional,kang2018self,ying2018sequential,li2020time}.
An addition benefit is that attention-based recommendation provides more interpretation, i.e., explain why the candidate items are recommended, therefore improving the users' satisfaction and the persuasiveness of RS.
In \cite{chen2017attentive}, an attention mechanism involving component- and item-level attention modules, is introduced in CF for multimedia recommendation.
Attentional Factorization Machines \cite{xiao2017attentional} learn the importance of each feature interaction via a neural attention network for context-aware recommendation.
In NARRE \cite{chen2018neural}, the attention mechanism was introduced to apply deep textual modeling on reviews for recommendation.
SHAN \cite{ying2018sequential} applies a two-layer attention module to obtain the long- and short-term preferences of a user in interaction sequences.
AttRec \cite{zhang2019next} combines self-attention mechanism and metric learning to learn better representations of preference.
TiSASRec \cite{li2020time} utilizes time-aware self-attention which learns the importance of temporal information for sequential recommendation.
In this paper, we use self-attention to extract the importance and contribution of temporal patterns to better model the logical expression in reasoning-based recommendation.

\section{Proposed Method} \label{sec_proposed}
In this section, we first give the problem formulation of recommendation throughout this paper and the definition of reasoning-based temporal recommendation, then we present the architecture of our \M~model, including each part of model architecture.
Finally, we provide the loss function of model training.

\begin{figure*} [!ht]
	\centering
	\includegraphics[width=\linewidth]{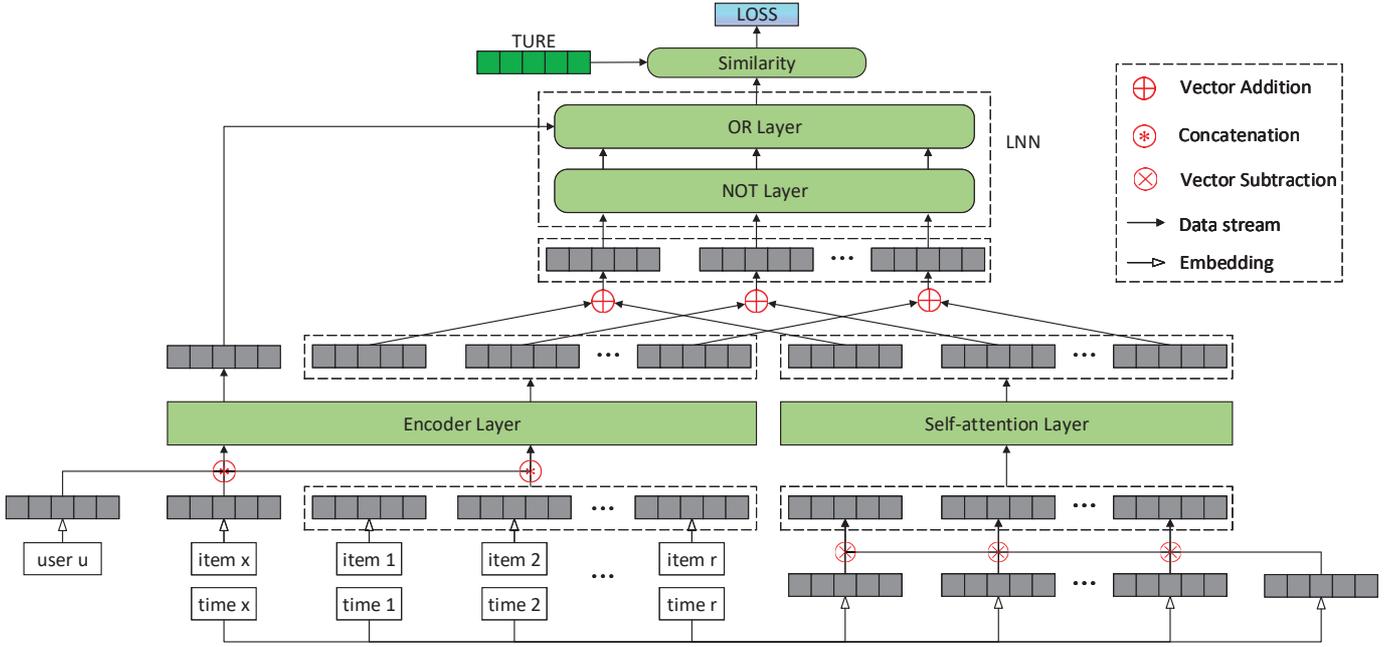}
	\caption{The overall architecture of \M.}
	\label{fig:architecture}
\end{figure*}

\subsection{Problem Formulation}
Let $\Uset$, $\Vset$ and $\Tset$ denote the sets of the users, items and timestamps respectively, $|\Uset|$, $|\Vset|$ and $|\Tset|$ denote the total number of users, items and timestamps respectively.
In order to improve the performance of reasoning-based recommendation, we focus on enhancing the capacity of logical reasoning by incorporating temporal information.
For each user $u \in \Uset$, his/her historical events of interaction are organized chronologically as $\mathcal{E}_u = \{ e_u^ {v_1,t_1}, e_u^ {v_2,t_2}, \cdots,  e_u^ {v_r ,t_r }\}$, where $r$ is the total number of items in the historical interactions, $e_u^ {v_i ,t_i }$ means that the user $u$ interacted with item $v_i$ at time $t_i$, where $v_i \in \Vset$ and $t_1 \leq t_2 \leq \cdots \leq t_r$ holds.

The task of temporal recommendation is to decide whether the next item $v_x$ at timestamp $t_x$ is recommended to the user $u$, according to his/her historical behaviors of interaction $\mathcal{E}_u$.
Hence, together with temporal information, the procedure of reasoning-based recommendation based on historical interactions can be formally defined to:
\begin{equation} \label{eq:lnn1}
	\left( e_u^{v_1,t_1} \wedge  e_u^{v_2,t_2} \wedge \cdots \wedge e_u^{v_r ,t_r} \right) \to e_u^ {v_x,t_x},
\end{equation}
which could be further rewritten as
\begin{equation} \label{eq:lnn2}
	\left( \neg e_u^{v_1,t_1} \vee \neg e_u^{v_2,t_2}  \vee \cdots \vee \neg e_u^{v_r ,t_r } \right) \vee e_u^ {v_x,t_x}
\end{equation}
according to De Morgan's Law.

For a fixed timestamp $t_x$, as to decide whether the candidate item $v_x$ is to be recommended or not, the learning procedure of temporal recommendation model would concentrate on the relative time, i.e., to learn the impact of relative time on the recommendation of candidate item.
Hence, to enhance the influence of temporal information on recommendation of the candidate item $v_x$, we employ self-attention mechanism to extract the importance of relative time.

\subsection{Extension to Explicit Feedback}
The procedure of reasoning-based recommendation in Eq.\eqref{eq:lnn1} considers the implicit feedback of user, i.e., each historical event $e_u^{v_i, t_i}$ only reveals the interaction between user $u$ and item $v_i$, but no information about whether $u$ likes or dislikes $v_i$.
In contrast, explicit feedbacks including ratings, add-to-favorite, like and dislike, provide more informative tendency of user towards the interacted items.
Consequently, leveraging explicit feedbacks in recommendation is beneficial to enhance the users' experience and improve the recommendation accuracy.

When extending it to reasoning-based recommendation, inspired by NCR \cite{chen2021neural}, we define $e_u^{v,t}$ and $\neg e_u^{v_i,t_i}$ as positive feedback and negative feedback from user $u$ on an interacted item $v$ at timestamp $t$, respectively.
Suppose that the historical events of interaction of user $u$ $\mathcal{E}_u = \{ e_u^ {v_1,t_1}, e_u^ {v_2,t_2}, \cdots,  \neg e_u^ {v_r ,t_r }\}$ describes the attitude of user $u$ towards interacted items $v_i$ at timestamp $t_i$.
Whether or not to recommend the candidate item $v_x$ at time $t_x$ could be expressed as the following logical statement:
\begin{equation} \label{eq:lnn_explicit1}
	\left( e_u^ {v_1,t_1} \wedge  e_u^{v_2,t_2} \wedge \cdots \wedge \neg e_u^{v_r ,t_r} \right) \to e_u^ {v_x,t_x}.
\end{equation}
Eq.\eqref{eq:lnn_explicit1} could be written equivalently as:
\begin{equation} \label{eq:lnn_explicit2}
	\left( \neg e_u^ {v_1,t_1} \vee \neg e_u^{v_2,t_2}  \vee \cdots \vee \neg \neg e_u^{v_r ,t_r } \right) \vee e_u^ {v_x,t_x},
\end{equation}
where the double negation is kept on $e_u^{v_r ,t_r}$ to enhance the expression of negation module of neural networks \cite{chen2021neural}.
Note that the negative feedbacks could be on more items.

\subsection{Network Architecture}
In this subsection, we present \M, a Time-aware Self-Attention with Neural Collaborative Reasoning based recommendation model.
The overall architecture of our model is shown in Fig. \ref{fig:architecture}.

First of all, given a user $u$, his/her interacted items $v_i$ at timestamp $t_i$ ($i=1,\cdots,r$) and candidate item $v_x$ at current timestamp $t_x$, the user $u$, items $\{v_1,\cdots, v_r, v_x \}$ and timestamps $\{t_1, \cdots, t_r, t_x\}$ are embedded to dense vectors in a differential space respectively.
To highlight the impact of temporal information on reasoning-based recommendation, we consider the relative time to represent the temporal information by subtracting the vectors of historical timestamps from the vector of current timestamp, which are then fed into the self-attention layer to capture the importance of temporal patterns.

On the other hand, the embedded vector of user and each embedded vectors of item are concatenated, which are further fed into an encode layer to encode the historical events.
The encoded historical events and the corresponding historical timestamps outputted by self-attention layer are simply added to represent the interactive events with temporal information, which are then fed into LNN layer, together with the concatenation of the embedded vector of user and that of candidate item.
At last, the similarity between the output vector of LNN layer and a constant true vector can be calculated for training the parameters of the model.

In the rest of this subsection, we introduce each part of \M~model in details.

\vspace{0.7em}
\noindent  \textbf{Embedding Layer:}
Due to the limited representation capacity of the original user IDs, item IDs and timestamps (i.e., one-hot representations) which are sparse and high dimensional, we employ three embedding layers to embed the users, items IDs, and the timestamp respectively, into three continuous low-dimensional spaces.
Each embedding layer is implemented by fully connected network to map the sparse inputs to dense vectors, which endow the inputs informative representation instead of discrete index.
Formally, given the sparse input features of user $\tilde \uv$, item $\tilde \vv$ and timestamp $\tilde \tv$, we can obtain their dense embeddings $\uv \in \R^d$, $\vv \in \R^d$ and $\tv \in \R^d$ via
\begin{equation}\label{eq:embed}
  \uv = \mathbf{E}_u^T \tilde{\uv}, ~~ \vv = \mathbf{E}_v^T \tilde{\vv},~~ \tv = \mathbf{E}_t^T \tilde{\tv}
\end{equation}
where $\mathbf{E}_u \in \R^{|\Uset| \times d}$, $\mathbf{E}_v \in \R^{|\Vset| \times d}$ and $\mathbf{E}_t\in \R^{|\Tset| \times d}$ are the embedding matrics for user features, item features and temporal features respectively, and $d$ is the dimension of the latent embedding space.

By means of the representation leaning technique, the latent embedding-based model is proven to be grate potential in capturing collaborative information.
Note that the $i$-th column of each embedding matrix represents the $i$-th sparse input feature, due to the identical mapping of one-hot representation matrix.
In practice, the operation of embedding is often implemented by look-up table retrieve, in which the embeddings are stored and retrieved using indices of input features.

\vspace{0.7em}
\noindent  \textbf{Self-attention Layer:}
As discussed in previous section, reasoning-based recommendation model incorporated with temporal information would tend to be self-attention.
Learning of reasoning-based recommendation model would gradually propagate to concentrate on relevant temporal patterns and inhibit irrelevances.
In order to enhance the impact of temporal information and to guide the learning of model to be self-attentive, we employ self-attention mechanism to consider different timestamps between two interactions in previous behaviours.
The self-attention automatically learns to seek the attentive weights of each timestamp: higher (lower) weights imply that the corresponding interactions are useful (useless) to candidate item.
It computes the dependence of each timestamp in previous interactions to build deeper representation of temporal patterns in reasoning-based recommendation.

To highlight the impact of temporal information on reasoning-based recommendation, we consider the relative time to capture temporal information.
Formally, given the embedding of timestamps $\tv_i (i=1,\cdots, r)$ in historical interactions and current timestamp $\tv_x$, the relative time is calculated by $\tv_i = \tv_x - \tv_i$.
Note that we overuse $\tv_i$ here, since the relative time can easily recover the timestamp according to current timestamp $\tv_x$.

Given a sequence of relative time in historical interactions $\{\tv_1,\tv_2,\cdots,\tv_r\}$, the self-attention mechanism computes a corresponding sequence of ``highlighted'' temporal information $\{\lv_1, \lv_2,\cdots,\lv_r\}$.
Each output element in the highlighted sequence $\lv_i$, is computed as a weighted sum of linearly transformed embeddings of centralized relative times, formally,
\begin{equation}
	\lv_i = \sum_{j=1}^{r} w_{ij}\left( \W^V \tv_j  \right) ,
\end{equation}
where $\W^V \in \R^{d \times d}$ is the trainable transformation matrix for value.
Each weight coefficient $w_{ij}$ is computed using a softmax function:
\begin{equation}
	w_{ij} = \frac{\exp(\alpha_{ij})} {\sum_{k=1}^{r}\exp(\alpha_{ik})},
\end{equation}
where $\alpha_{ij}$ is the contextual relevance of relative timestamps $\tv_i$ and $\tv_j$, and the relevance is measured via the dot product:
\begin{equation}
	\alpha_{ij}=\frac{ \left( \W^Q \tv_i \right) \left( \W^K \tv_j \right)^T }{\sqrt{d}},
\end{equation}
where $\W^Q \in \R^{d \times d}$ and $\W^K \in \R^{d \times d}$ are the trainable projection matrices to distill useful time information for propagation of logical reasoning.
The denominator $\sqrt{d}$ is the limit coefficient which prevents the value of the inner product above from being too large, especially when the dimension is high.

\vspace{0.7em}
\noindent  \textbf{Encoder Layer:}
In order to model user-item interactions, CF-based recommendation models are widely employed, where the preference of user is commonly modeled with a linear kernel, i.e., the inner product of user and item embeddings.
However, trivially combining the multiplication of latent embeddings linearly, the inner product may not be sufficient to capture the complex structure of user-item interactions well, which significantly degrades the performance of recommendation.

To model the non-linear interactions between user and item, instead, MLP is commonly to fuse the dense embeddings of user and item, which is standard practice to learn deeper representation in industrial RS.
We adopt a simple neural network to encode the non-linearity of interaction between user embedding $\uv$ and item embedding $\vv$ into event vector $\boldsymbol{e}^v_u$, which we regard as non-linear fusion.
The neural network takes the concatenation of user and item embeddings as input, and contains two linear transformations with a ReLU activation in between, formally,
\begin{equation}
	\boldsymbol{e}^v_u=\W_2\phi (\W_1 (\uv \circledast \vv) + \bv_1) + \bv_2
\end{equation}
where $\uv, \vv \in \mathbb{R}^d$ are user and item latent embedding vectors in $d$-dimensional space; $\circledast$ is concatenation operation between user and item embeddings; $\W_1 \in \R^{2d \times d }$, $\W _2 \in \R^{d \times d}$ are weight matrices and $\bv_1, \bv_2 \in \R^d$ are bias terms;
$\boldsymbol{e}^v_u \in \R^d$ represents the encoded event embedding, $\phi (\cdot ) = \max(0,x)$ is the rectified linear unit (ReLU) activation function.

Furthermore, in order to fuse temporal patterns outputted by self-attention layer into user-item interactions, we encode each temporal event vector $\boldsymbol{e}_{u}^{v_i,t_i}$ by simply adding self-attentive time information $\lv_i$ accordingly to event vector $\boldsymbol{e}_{u}^{v_i}$, i.e., $\boldsymbol{e}_{u}^{v_i,v_i} = \boldsymbol{e}_{u}^{v_i} + \lv_i$, ($i=1,\cdots,r$).
It is worth mentioning that the temporal event vector $\boldsymbol{e}_{u}^{v_x,t_x}$ degenerates to the event vector $\boldsymbol{e}_{u}^{v_x}$.
Since that the relative time of current timestamp is zero, the self-attentive time information $\lv_x$ is the arithmetic average of relative time embeddings $\tv_i$, i.e., $\lv_x = \frac{1}{r} \sum_{i=1}^{r} \W^V \tv_i$, leading to no attention of $\lv_x$ on previous temporal information.
Hence, it is non-trivial to fuse event embedding and highlighted temporal embedding by applying addition between them.

\vspace{0.7em}
\noindent  \textbf{LNN Layer:}
In this paper, we apply neural logical reasoning to simulate the procedure of recommendation, where the self-attentive temporal information are incorporated to highlight the importance of temporal patterns of reasoning-based recommendation.
To this end, we have so far obtained the temporal event vector by encoding the highlighted temporal embedding obtained by self-attention, and the fused event vector which represents the non-linearity of user-item interaction.
Based upon the fusion of event embeddings and highlighted temporal embeddings, the next step is to utilize logical neural network to predict the logical expression in Eq.\eqref{eq:lnn2}, which is now represented as the logical aggregation of temporal event embeddings:
\begin{equation}\label{eq:lnn_vec}
  \left( \neg \boldsymbol{e}_{u}^{v_1,t_1} \vee \neg \boldsymbol{e}_{u}^{v_2,t_2} \vee \cdots \vee \neg \boldsymbol{e}_{u}^{v_r,t_r} \right) \vee \boldsymbol{e}_{u}^{v_x,t_x}.
\end{equation}

In order to carry out the prediction of logical expression, NCR \cite{chen2021neural} employs LNN layer to perform logical reasoning based on the embeddings of interaction events.
We follow this pipeline, but the difference is that the prediction of logical reasoning is based on the temporal event vectors as shown in Eq.\eqref{eq:lnn_vec}, which fuses not only interaction events but also highlighted temporal information.
As a result, the fusion of self-attentive temporal information into interaction events provides deeper representation of user's preference in reasoning-based recommendation.

According to the logical expression in Eq.\eqref{eq:lnn_vec}, we firstly calculate the negated temporal event $\neg \boldsymbol{e}_{u}^{v_i,t_i}$ by feeding each input temporal event $\boldsymbol{e}_{u}^{v_i,t_i}$ into NOT layer:
\begin{equation}
	\neg \boldsymbol{e}_{u}^{v_i,t_i} = \text{NOT} \left(\boldsymbol{e}_{u}^{v_i,t_i} \right) , ~~~ \forall i \in \{1,2,...,r\}
\end{equation}
After that, along with above negated temporal events $\neg \boldsymbol{e}_{u}^{v_i,t_i}$, the candidate temporal event $\boldsymbol{e}_{u}^{v_x,t_x}$ is fed into the OR layer, of which the output generates the final embedding of the logical expression in Eq.\eqref{eq:lnn_vec} as follow
\begin{equation}
	\textbf{Exp}=\text{OR}(\neg \boldsymbol{e}_{u}^{v_1,t_1} , \neg \boldsymbol{e}_{u}^{v_2,t_2} , \cdots , \neg \boldsymbol{e}_{u}^{v_r,t_r} , \boldsymbol{e}_{u}^{v_x,t_x})
\end{equation}
where \textbf{Exp} is the vector representation of the logical expression, and the operator $\text{OR}$ is the logical disjunction of all input temporal event vectors.
This can be implemented by recurrently calling the basic binary model $\text{OR}(\cdot,\cdot)$ as shown at line 5 in following procedure:
\begin{center}
  \shadowbox{
  \parbox{0.45\textwidth}{
    \textbf{Procedure}: $\text{LNN}(\boldsymbol{e}_{u}^{v_1,t_1} , \boldsymbol{e}_{u}^{v_2,t_2} , \cdots ,  \boldsymbol{e}_{u}^{v_r,t_r} , \boldsymbol{e}_{u}^{v_x,t_x})$ \\
    1. \textbf{Input}: $\boldsymbol{e}_{u}^{v_1,t_1} , \boldsymbol{e}_{u}^{v_2,t_2} , \cdots ,  \boldsymbol{e}_{u}^{v_r,t_r} , \boldsymbol{e}_{u}^{v_x,t_x}$. \\
    2. \textbf{Initialization}: $\textbf{Exp}=\boldsymbol{e}_{u}^{v_x,t_x}$, $\mathcal{I}=\{ 1,\cdots, r \}$.\\
    3. \textbf{while} $\mathcal{I}$ is not empty: \\
    4. ~~Randomly draw $i$ from $\mathcal{I}$ without replacement.\\
    5. ~~$\textbf{Exp} \leftarrow \text{OR}\left(\textbf{Exp}, \text{NOT}\left( \boldsymbol{e}_{u}^{v_i,t_i} \right) \right)$.\\
    6. \textbf{end while} \\
    7. \textbf{Return}: $\textbf{Exp}$
  }
}
\end{center}

\subsection{Loss Function}
After obtaining the vector representation of logical expression in Eq.\eqref{eq:lnn_vec} outputted by LNN layer, the model should determine whether $\textbf{Exp}$ represents true or false.
To this end, similar to NCR \cite{chen2021neural}, the vector representation is examined in the logical space by computing the distance between the vector $\textbf{Exp}$ and the anchor vector ($\T$) which represents the logical truth and is randomly initialized but fixed during model training and inference.
We employ the cosine similarity to measure this distance:
\begin{equation} \label{eq:sim}
	\textsl{Sim}(\textbf{Exp}, \T) = \frac{ \left \langle \textbf{Exp}, \T \right \rangle}{\left \| \textbf{Exp}  \right \|  \left \| \T  \right \|}.
\end{equation}

The goal of our \M~model is to determine whether the candidate item is to be recommended to user or not, given the sequence of interactions including the pairs of interacted items and corresponding timestamps.
To achieve this goal, we employ the pair-wise learning algorithm \cite{rendle2012bpr} for model training, of which the objectives usually concentrate on users' preference with respect to the pairs of items.
In pair-wise learning algorithm for recommendation system, the objective is to jointly encourage the scores of positive samples and suppress that of counter samples, given the historical events of interactions.

For a user $u$, according to his/her interactions, the set of items $\Vset$ can be divided to two disjoint set $\Vset_u^+$ and $\Vset_u^-$, where $\Vset_u^+$ denotes the set of items that the user $u$ interacted with, while $\Vset_u^-$ represents those of that the user $u$ did not interacted with, $\Vset_u^+ \cap \Vset_u^- = \emptyset$ and $\Vset_u^+ \cup \Vset_u^- = \Vset$ hold.
Let the pairs of items and timestamps $\{ (v_1,t_1), (v_2,t_2),\cdots, (v_r,t_r)\}$ be the historical interactions of user $u$.
We sample item $v_x \in \Vset_u^+$ with timestamp $t_x$ to construct the positive sample that user $u$ interacted with, and item $v_y \in \Vset_u^-$ with timestamp $t_y$ to construct the negative sample that user $u$ did not interact with.

The logical expressions of positive sample and counter sample recommendation could be modelled as
\begin{equation} \label{eq:exp_pos_neg}
	\begin{split}
		\boldsymbol{E}^+_{u,x} &= \left( \neg \boldsymbol{e}_{u}^{v_1,t_1} \vee \neg \boldsymbol{e}_{u}^{v_2,t_2} \vee \cdots \vee \neg \boldsymbol{e}_{u}^{v_r,t_r} \right) \vee \boldsymbol{e}_{u}^{v_x,t_x}  \\
		\boldsymbol{E}^-_{u,y} &= \left( \neg \boldsymbol{e}_{u}^{v_1,t_1} \vee \neg \boldsymbol{e}_{u}^{v_2,t_2} \vee \cdots \vee \neg \boldsymbol{e}_{u}^{v_r,t_r} \right) \vee \boldsymbol{e}_{u}^{v_y,t_y}
	\end{split}
\end{equation}
where $\boldsymbol{E}^+_{u,x}$ and $\boldsymbol{E}^-_{u,y}$ are the expressions of the ground-truth interaction and the negative sampled interaction, respectively.
Notably, both above logical expressions fuse the temporal information into interactions.
Consequently, the truth evaluations of these expressions depend on not only the interacted items but also the timestamp, e.g., $\boldsymbol{E}^+_{u,x}$ may be false if the timestamp is not $t_x$, while $\boldsymbol{E}^-_{u,y}$ would be true if the timestamp is not $t_y$.

To evaluate the truth of expressions in Eq.\eqref{eq:exp_pos_neg}, we define $s^+_{u,x}$ and $s^-_{u,y}$ to be the prediction scores for an positive sample of item $v_x$ at timestamp $t_x$ and a negative sample of item $v_y$ at timestamp $t_y$, respectively.
Simulated by LNN layer, the expressions in Eq.\eqref{eq:exp_pos_neg} is further evaluated in term of similarity via Eq.\eqref{eq:sim}, giving the prediction scores below:
\begin{equation} \nonumber 
	\begin{split}
		s^+_{u,x} &= \textsl{Sim}(\text{LNN}(\boldsymbol{e}_{u}^{v_1,t_1} , \boldsymbol{e}_{u}^{v_2,t_2} , \cdots ,  \boldsymbol{e}_{u}^{v_r,t_r} , \boldsymbol{e}_{u}^{v_x,t_x}), \T)
		\\
		s^-_{u,y} &= \textsl{Sim}(\text{LNN}(\boldsymbol{e}_{u}^{v_1,t_1} , \boldsymbol{e}_{u}^{v_2,t_2} , \cdots ,  \boldsymbol{e}_{u}^{v_r,t_r} , \boldsymbol{e}_{u}^{v_y,t_y}), \T)
	\end{split}.
\end{equation}

In order to guarantee the \M~to encourage the prediction score of positive sample and suppress that of negative sample, i.e., to jointly maximize $s^+_{u,x}$ and minimize $s^-_{u,y}$, following NCR \cite{chen2021neural}, we also construct the difference of these two prediction scores as $s_u^{x,y} = \beta \cdot (s^+_{u,x} - s^-_{u,y})$, where $\beta$ is the amplification factor.
In this scenario, the objective of model training turns to be maximization of this difference.
The difference $s_u^{x,y}$ is further employed to derive the objective function of pair-wise learning, formally,
\begin{equation}
	\mathcal{L}_{\text{pair-wise}} = - \sum_{u \in \Uset} \sum_{v_x \in \Vset^+_u} \sum_{v_y \in \Vset^-_u} \ln \sigma (s_u^{x,y}) + \lambda_\Delta  \left\| \Delta \right\|^2_2,
\end{equation}
where $\Delta$ is all parameters of our model, including the embedding matrices of user, item and timestamp, the parameters of self-attention layer, the parameters of the encoding layer, and the parameters of LNN layer;
$\lambda_\Delta$ is the coefficient of $\ell_2$-norm regularization term which prevents the weights from being too large;
$\sigma(\cdot)$ is the sigmoid function: $\sigma(x) = \frac{1}{1+e^{-x}}$.

Note that the LNN layer which simulates the logical expression in the space of event vectors, is a plain neural architecture expected to really realize logical reasoning.
To this end, we also employ the idea of logical regulation in NCR, in which logical regularizer was introduced to constrain the logical neural modules.
Integrating the logical regularizer into the objective function in pair-wise learning gives the final loss function of our model training:
\begin{equation} \label{eq:final_loss}
	\mathcal{L} = \mathcal{L}_{\text{pair-wise}} + \lambda_r \mathcal{L}_{\text{logicReg}},
\end{equation}
where $\lambda_r$ is the coefficient for the logic regularizer, and $\mathcal{L}_{\text{logicReg}}$ is the logic regularizer representing several logical laws, such as negation and double negation in NOT module, identity, annihilator, idempotence and complementation in OR module, etc.\cite{chen2021neural}

It is worth mentioning that in terms of the logical regulation, the logical regularizer in NCR \cite{chen2021neural} involves the logical expression of the fused embeddings of non-temporal user-item interactions, while our \M~model deals with that of self-attentive temporal interactions of users and items, leading to deeper representation of logical constrains for reasoning-based recommendation.

\section{Experiments} \label{sec_exp}
In order to validate the effectiveness of integrating self-attentive temporal information into reasoning-based recommendation, in this section, we conduct experiments on benchmark datasets to demonstrate the advantage of proposed \M~model over other state-of-the-art recommendation models, including  non-temporal reasoning-based model.
In particular, we answer the following research questions from experimental point of view:
\begin{itemize}
  \item \textbf{RQ1:} What’s the performance of our proposed \M~model compared to other state-of-the-art methods for recommendation systems?
  \item \textbf{RQ2:} Does the integration of temporal information make contribute to improvement of reasoning-based recommendation?
  Furthermore, for the representation of temporal information, which is the better choice for relative time and absolute time?
  \item \textbf{RQ3:} What is the impact of self-attention on the model performance when considering temporal patterns in reasoning-based recommendation?
  \item \textbf{RQ4:} What are the effects of hyper-parameters on the performance of \M~model, including the number of dimensions and that of training epoches?
  What's the recommendation performances of \M~model under the metrics with different ranks, e.g., HR@$k$ and NDCG@$k$ with different rank $k$.
\end{itemize}

\subsection{Experimental Setup}
\noindent  \textbf{Datasets:}
We evaluate our proposed \M~model and compare it with several baselines by conducting experiments on three public real-world datasets: MovieLens 100K\cite{harper2015movielens},  Amazon Movies\text{\&}TV and Amazon Electronics \cite{mcauley2015image}. The statistics of these three datasets are summarized in Table \ref{tab_dataset}.

\begin{table} [!h]
    \caption{Statistical summary of datasets.}
    \label{tab_dataset}
	\centering
	\begin{tabular}{ccccc}
		\toprule
		Dataset               &\#Users         &\#Items          &\#Interaction     & Density   \\
		\midrule
		MovieLens100K        & 943            & 1,682           &100,000           & 6.3\%      \\
		Movies\&TV       & 123,961        & 50,053          &1,697,533         & 0.027\%    \\
		Electronics    & 192,404        & 63,002          &1,689,188         & 0.014\%  \\
		\bottomrule
	\end{tabular}
\end{table}

MovieLens100K dataset was collected by the GroupLens Research Project at the University of Minnesota, and contains 100,000 movie ratings ranging from 1 to 5 from 943 users on 1,682 movies, and each user has rated at least 20 movies.
Movies\&TV and Electronics which are the Amazon e-commerce datasets, contain product reviews and metadata spanning from May 1996 to July 2014, including users, items and rating information.
Compared with MovieLens100K, both these two datasets are relatively sparse and cover about two-million data of ratings.
The above mentioned datasets are widely used in the recommendation systems literatures for model evaluation.

We follow the same procedure of data preprocessing from \cite{chen2021neural}.
For reasoning-based recommendation with implicit feedback, we consider the interactions between user and item without ratings to build the logical expression in Eq.\eqref{eq:lnn1}.
While reasoning with the explicit feedback, for all datasets, we transform the ratings which range from 1 to 5, to 0 and 1.
In particular, we treat the ratings of 4 and 5 as positive feedback which are set to 1, and those of 1 to 3 as negative feedback which are set to 0.
For each user, we sort his/her historical events of interaction chronologically in terms of timestamp.
Given $e_u^{v_x,t_x}$ the event of interaction of user $u$ and an item $v_x$ at timestamp $t_x$, we select the most recent 5 historical interactions to construct the logical expression, in Eq.\eqref{eq:lnn1} with implicit feedback or Eq.\eqref{eq:lnn_explicit1} with explicit feedback.

For each user, events from earliest 5 interactions and those with less than 5 interactions are directly put into training set.
For partitioning of each dataset, we conduct leave-one-out operation on the sequence of historical interactions for each user to create test set and validation set.
The most recent events and the second most recent events of interaction are assigned to the test set and validation set respectively, while all remaining events are assigned to training set.
Notice that training set and validation set would be leveraged to build logical expression during test phase.

\vspace{0.5em}
\noindent  \textbf{Baselines:}
We compare the proposed model with the following baseline algorithms, including shallow and deep models:
\begin{itemize}
  \item \textbf{BPR-MF}\cite{rendle2012bpr}. The Bayesian personalized ranking is a classic recommendation model based on implicit feedback, which is optimized by a pair-wise learning. We choose Biased Matrix Factorization \cite{koren2009matrix} as the internal predictor.
  \item \textbf{SVD++}\cite{koren2008factorization}. It extends Singular Value Decomposition (SVD) with neighborhood by considering user history interactions, where the item-item similarity is modeled by adding a second set of item factors.
  \item \textbf{DMF}\cite{xue2017deep}. Deep Matrix Factorization employs multi-layer perceptrons for user and item matching, in which an inner product of user and item embeddings is adopted to calculate the interactions, and is optimized on a point-wise loss function.
  \item \textbf{NCF}\cite{he2017neural}. The state-of-the-art method of neural network-based collaborative filtering algorithm for item recommendation, which incorporates the hidden layers of generalized matrix factorization and multi-layer perceptrons to learn the non-linear interactions between users and items.
  \item \textbf{GRU4Rec}\cite{hidasi2015session}. It is a session-based recommendation model which employs Recurrent Neural Network (RNN) structured GRU with ranking-based loss to model user preferences for sequential recommendation.
  \item \textbf{STAMP}\cite{liu2018stamp}. The Short-Term Attention/Memory Priority model, which fully employs attention mechanism to capture the user's long- and short-term preference from previous clicks in a session.
      It uses attention layers to replace all RNN encoders to consider recent interactions in a sequence.
  \item \textbf{NCR}\cite{chen2021neural}. It models recommendation as a reasoning task by integrating logical structure and neural networks, where neural networks capture similarity patterns in data while logical reasoning facilitates cognitive reasoning for decision making for recommendation.
\end{itemize}

In addition, we test three versions of our model to prove the effectiveness of integrating temporal information and self-attention in reasoning-based recommendation.
We consider absolute timestamp and relative time to represent the temporal information.
\begin{itemize}
	\item \textbf{\M-A w/o SA} Absolute Time-aware Self-Attention with Neural Collaborative Reasoning without Self-Attention, which use absolute timestamps as temporal information for model learning.
	\item \textbf{\M-R w/o SA} Relative Time-aware Self-Attention with Neural Collaborative Reasoning without Self-Attention, which use relative timestamps as temporal information for model learning.
	\item \textbf{\M} Time-aware Self-Attention with Neural Collaborative Reasoning, which use self-attentive output of relative timestamps as temporal information for model learning.
\end{itemize}

\vspace{0.5em}
\noindent  \textbf{Metrics:}
To evaluate the performance of our proposed method and compare it with other baselines for recommendation problem, we employ two widely used ranking based metrics, namely Normalized Discounted Cumulative Gain at rank $k$ (NDCG@$k$) and Hit Ratio at rank $k$ (HR@$k$).
For these two metrics, the larger values of both indicates better performance of model.

NDCG is a position-aware metric which considers the position and assigns higher score to higher position.
It calculates the ranking quality by normalizing the Discounted Cumulative Gain (DCG) measure which is a weighted sum of the degree of relevancy of the ranking items, formally:
\begin{equation}
	\text{NDCG@}k = \frac{1}{Z} \text{DCG@}k = \frac{1}{Z} \sum_{i=1}^{k} \frac{2^{r_i} - 1}{\log_2(i+1)}
\end{equation}
where $r_i \in {1,0}$ is the graded relevance of the result at position $i$, and $Z$ is a normalization factor which is the ideal DCG@$k$ defined as $\max_{r_i} \sum_{i=1}^{k} \frac{2^{r_i} - 1}{\log_2(i+1)}$.

HR measures the proportion of the number that the ground-truth items that appears in the recommended list of length $k$, which is formulated as follows:
\begin{equation}
	\text{HR@}k = \frac{1}{|\Uset|} \sum_{i=1}^{|\Uset|} \frac{\# hit_i\text{@}k}{\#I(u_i)}
\end{equation}
where $\# hit_i\text{@}k$ is the number of hits for user $u_i$ and $\#I(u_i)$ denotes the total number of interactions of user $u_i$.

In the experiments, we employ the popular setting of $k=5,10,20$ for validation, and the result of both metrics are averaged over all users to reduce the effect of random oscillation.
To avoid heavy computational burden, for each user, $100$ negative samples are randomly drawn and mixed with the ground-truth (i.e., positive sample) in the ranking process.

\vspace{0.5em}
\noindent  \textbf{Implementation details:}
The parameters for all baselines were initialized as in the corresponding papers, and carefully tuned to obtain the optimal performance.
The implementation details of the proposed \M~are provided as follows.
The dimensions of users, item, and event embedding latent vectors are determined by grid search in the range of \{$20,40,60,80,100$\}.
The weight of logical regularizer $\lambda_r$ in Eq.\eqref{eq:final_loss} is set to $0.01$.
For model training, the batch size of training examples is set to 128, and the maximum of training epoch is determined by grid search in the range of \{$50,60,80,100,200,500,1000$\}.
Both $\ell_2$ regularization and dropout are employed to prevent model fitting.
The weight of $\ell_2$ regularization and dropout ratio are set to $10^{-4}$ and $0.2$, respectively.
All of the parameters are first initialized according to Gaussian distribution $\mathcal{N}(0,0.01)$, and then updated with Adam optimizer.
The learning rate of Adam is set to $10^{-3}$.
All experiments on benchmark datasets are implemented with PyTorch, on a 64 core Intel Xeon Gold 6226R CPU @ 2.90GHz, 256 GB memory and a Nvidia Duadro RTX 8000 GPU.

\begin{table*}[!ht]
	\caption{Results of recommendation performance on three datasets, and all of the numbers in the table are percentages with \% omitted. We use underwave(\uwave{number}) to mark the best result among all baselines. We use underline(\uline{number}) to show the best result between Absolute and Relative Time-aware Self-Attention with Neural Collaborative Reasoning without self-attention. We use bold font to denote the best result among all baselines and three versions of proposed model. $\text{Improvement}^1$ implies that the proposed \M~improvement over NCR; $\text{Improvement}^2$ shows that the improvement of \M-A w/o SA over NCR, where -- denotes no improvement; $\text{Improvement}^3$ demonstrates the improvement of \M-R w/o SA over NCR; $\text{Improvement}^4$ manifests the improvement of \M-R w/o SA over \M-A w/o SA; $\text{Improvement}^5$ indicates the improvement of \M~over \M-R w/o SA.}
	\setlength\tabcolsep{2pt}
	\label{tab_results}
	\centering
	\begin{tabular}{ccccccccccccccc}
		\toprule
		& \multicolumn{4}{c}{MovieLens100K}    &    & \multicolumn{4}{c}{Movies\&TV} & & \multicolumn{4}{c}{Electronics}     \\
		\cline{2-5} \cline{7-10} \cline{12-15}
		& N@5 & N@10 & HR@5   & HR@10 &  & N@5 & N@10 & HR@5   & HR@10 & & N@5  & N@10 & HR@5   & HR@10   \\
		\midrule
		BPR-MF       & 0.3024 & 0.3659  & 0.4501 & 0.6486 & & 0.3962 & 0.4392  & 0.5346 & 0.6676 & & 0.3092  & 0.3472  & 0.4179 & 0.5354  \\
		SVD++        & 0.3087 & 0.3685  & 0.4586 & 0.6433 & & 0.3918 & 0.4335  & 0.5224 & 0.6512 & & 0.2775  & 0.3172  & 0.3848 & 0.5077  \\
		DMF          & 0.3023 & 0.3661  & 0.4480 & 0.6450 & & 0.4006 & 0.4455  & 0.5455 & 0.6843 & & 0.2775  & 0.3143  & 0.3783 & 0.4922  \\
		NCF          & 0.3002 & 0.3592  & 0.4490 & 0.6316 & & 0.3791 & 0.4211  & 0.5134 & 0.6429 & & 0.3026  & 0.3358  & 0.4031 & 0.5123  \\
		GRU4Rec      & 0.3564 & 0.4122  & 0.5134 & 0.6856 & & 0.4038 & 0.4459  & 0.5287 & 0.6688 & & 0.3154  & 0.3551  & 0.4284 & 0.5511  \\
		STAMP        & 0.3560 & 0.4070  & 0.5159 & 0.6730 & & 0.3935 & 0.4366  & 0.5246 & 0.6577 & & 0.3095  & 0.3489  & 0.4196 & 0.5430  \\
		NCR          & \uwave{0.3760} & \uwave{0.4240}  & \uwave{0.5456} & \uwave{0.6943} & & \uwave{0.4255} & \uwave{0.4670}  & \uwave{0.5611} & \uwave{0.6891} & & \uwave{0.3499}  & \uwave{0.3878}  & \uwave{0.4639} & \uwave{0.5812}  \\
		\hline
		\M-A w/o SA       & 0.3722 & 0.4289 & 0.5304 & 0.7053 & & 0.4344 & 0.4753 & 0.5701 & 0.6964 & & 0.3680 &	0.4076 & 0.4890 & 0.6114  \\
		\M-R w/o SA          & \underline{0.3913} & \underline{0.4469} & \underline{0.5557} & \underline{0.7267} & & \underline{0.4451} & \underline{0.4857} & \underline{0.5861} & \underline{0.7114} & & \underline{0.3761} & \underline{0.4158} & \underline{0.4970} & \underline{0.6195}  \\
		\M      &\textbf{0.4030} & \textbf{0.4560}  & \textbf{0.5654} & \textbf{0.7289} & & \textbf{0.4511} & \textbf{0.4911}  & \textbf{0.5911} & \textbf{0.7145} & & \textbf{0.3769}  & \textbf{0.4166}  & \textbf{0.4985} & \textbf{0.6212}  \\
		\hline
		\hline
		$\text{Improvement}^1$ & 7.18\% & 7.55\%  & 3.63\% & 4.98\% & & 6.02\% & 5.16\%  & 5.35\% & 3.69\% & & 7.72\%  & 7.43\%  & 7.46\% & 6.88\%  \\
		$\text{Improvement}^2$ & -- & 1.16\% & -- & 1.58\% & & 2.09\% & 1.78\% & 1.60\% & 1.06\% & & 5.17\% & 5.11\% & 5.41\% & 5.20\% \\
		$\text{Improvement}^3$ & 4.07\% & 5.40\% & 1.85\% & 4.67\% & & 4.61\% & 4.00\% & 4.46\% & 3.24\% & & 7.49\% & 7.22\% & 7.14\% & 6.59\% \\
		$\text{Improvement}^4$ & 5.13\% & 4.20\% & 4.77\% & 3.03\% & & 2.46\% & 2.19\% & 2.81\% & 2.15\% & & 2.20\% & 2.01\% & 1.64\% & 1.32\%  \\
		$\text{Improvement}^5$ & 2.99\% & 2.04\% & 1.75\% & 0.30\% & & 1.35\% & 1.11\% & 0.85\% & 0.44\% & & 0.21\% & 0.19\% & 0.30\% & 0.27\%  \\
		\bottomrule
	\end{tabular}
\end{table*}
\normalem

\subsection{Comparison of Overall Performance}
To answer \textbf{RQ1}, we evaluate the recommendation performance of our proposed \M~model and the baselines.
Table \ref{tab_results} reports the comparisons w.r.t. metrics NDCG(N) and HR at rank 5 and 10 on all of the datasets.
We use underwave to mark the best result among all baselines including matching-based recommendations and NCR (i.e., reasoning-based recommendation), and we use underline to highlight the best result between \M-A w/o SA and \M-R w/o SA, where the former incorporates absolute timestamp into NCR while the latter incorporates relative time into NCR.
Numbers with bold fonts indicate the best result of all recommendation methods, including the baselines, our proposed model \M~and other two versions of it.

We first analyze the recommendation results of the baselines.
As sequence/session recommendation models, GRU4Rec and STAMP mostly achieve best performance than the first four models upon all three datasets in terms of all metrics.
We can see that GRU4Rec and STAMP are the best among matching-based recommendation models.
On the other hand, NCR obtains the best performance among the baselines in terms of all metrics, which implies that the integration of perception learning and cognitive reasoning into recommendation task provides significant improvement on the ranking performance.
Hence, we refer NCR to the best result of baselines for comparisons in further experiments.

Next we evaluate our proposed \M~model against the baselines in terms of the ranking performance.
As shown in Table \ref{tab_results}, the numbers with bold fonts indicate that our proposed model \M~consistently and noticeably outperforms NCR, i.e., the best result among baselines on all of the datasets.
This shows that the incorporation of temporal information and self-attention mechanism into reasoning-based recommendation helps to improve the recommendation performance.
In particular, significant improvements against NCR are observed in Table \ref{tab_results}.
Compared with NCR, the ranking result of the proposed model \M~historically walks into $0.4$ in terms of N@5 and $0.7$ in terms of HR@10 on the MovieLens100K dataset, $0.7$ in terms of HR@10 on the Movies\&TV dataset, $0.4$ in terms of N@10 and $0.6$ in terms of HR10 on the Electronics dataset.
$\text{Improvement}^1$ row in Table \ref{tab_results} implies the percentage improvement of proposed \M~ model against NCR.
Our method \M~outperforms all baselines on all datasets and obtains at most $7.72\%$ NDCG@5, $7.75\%$ NDCG@10, $7.46\%$ HR@5 and $6.88\%$ HR@10 improvements against the best baseline recommendation model.
The significant improvement over reasoning-based recommendation may be due to the incorporation of temporal information and self-attention mechanism, which motivates us to further investigate the influences of temporal information and self-attention mechanism on reasoning-based recommendation model to answer \textbf{RQ2} and \textbf{RQ3}.

\subsection{Influence of Absolute and Relative Time Pattern}
There are two typical temporal patterns of users behavior in recommendation system, including absolute time pattern and relative time pattern, in which, the former is a point-wise variable that provides the context and auxiliary information to characterize the preference of user, while the latter is a pair-wise concept which represents temporal information as the time interval between each pair of interactions.
In order to investigate the pros and cons of these two temporal patterns in reasoning-based recommendation, we conduct ablation experiments on the three datasets mentioned above.

When considering absolute time pattern to be temporal context in reasoning-based recommendation, we extract absolute timestamps in each user-item interaction and fed them into LNN layer without self-attention layer in network architecture, leading to the \M-A w/o SA model.
In Table \ref{tab_results}, the eighth row shows the ranking performance on three datasets, and $\text{Improvement}^2$ row reports the improvement of \M-A w/o SA over NCR.
Compared with NCR, the \M-A w/o SA model obtains little even no improvement on the ranking performance.
Hence, leveraging absolute time pattern can not guarantee improvement to reasoning-based recommendation, sometimes would ruin the improvement of model performance.

On the other hand, we turn to integrate relative time pattern to reasoning-based recommendation.
In network architecture, we extract relative times, i.e., time intervals between previous interactions and current stage, and also fed them into LNN layer without self-attention layer, resulting in the \M-R w/o SA model.
In Table \ref{tab_results}, the ninth row shows the ranking performance of the \M-R w/o SA model on three datasets, and $\text{Improvement}^3$ row reports the improvement of \M-R w/o SA over NCR.
We can observe that the ranking performance of the \M-R w/o SA model is superior to those of baselines and the \M-A w/o SA model.
It is not surprising that relative time pattern probably provides more highlighted relevance between historical interactions and candidate item at current timestamp.
$\text{Improvement}^4$ row manifests the improvement of \M-R w/o SA model over of \M-A w/o SA model, which also confirms that relative time patterns are extremely important for reasoning-based recommendation.
Compared with the best baseline NCR, the integration of relative time patterns in reasoning-based recommendation shows significant superiority to model the temporal dynamics of user's preference, hence improving the effectiveness of logical reasoning in recommendation.

\subsection{Influence of Self-Attention}
As mentioned above, integration of relative time patterns in reasoning-based recommendation provides grate improvement of ranking performance, however the model of RS would tend to be self-attentive on temporal information.
Here we investigate the impact of self-attention mechanism on the ranking performance of reasoning-based recommendation incorporated with relative time patterns.
In order to answer \textbf{RQ3}, we mainly focus on discussing two variations of recommendation models, namely \M-R w/o SA and \M, where self-attention mechanism is (not) leveraged to distill informative temporal patterns and suppress irrelevance in \M~(\M-R w/o SA) model, respectively.

In Table \ref{tab_results}, $\text{Improvement}^5$ row indicates the improvement of proposed \M~over \M-R w/o SA.
It can be observed that employing self-attention mechanism to highlight the importance of temporal patterns helps further improve the ranking performance of recommendation.
Such observation demonstrates the effectiveness of self-attention to extract the importance of relative time patterns for reasoning-based recommendation.
It is interesting to see that the improvements of leveraging self-attention are slight compared with that of integrating relative time patterns.
The reason may be that during training, the weights of model with temporal information gradually concentrates on most relevance of temporal patterns, and the self-attention is further employed to encourage the model to distill informative patterns and suppress irrelevance.

\subsection{Ablation Study on Hyper-parameters}
In order to answer \textbf{RQ4}, in this section, we investigate the impacts of hyper-parameters via an ablation study.
The hyper-parameters include the number of dimensions of users, item, and event embedding latent vectors, and the number of training epoch.
We also investigate the ranking performance of proposed model \M~over the metrics HR and NDCG at different rank, i.e., HR@$k$ and NDCG@$k$.

\begin{figure}[!t]
	\centering
	\subfloat{\includegraphics[width=0.25\textwidth]{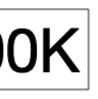} }\\
	\setcounter{subfigure}{0}
	\subfloat[HR@5]{\includegraphics[width=0.24\textwidth]{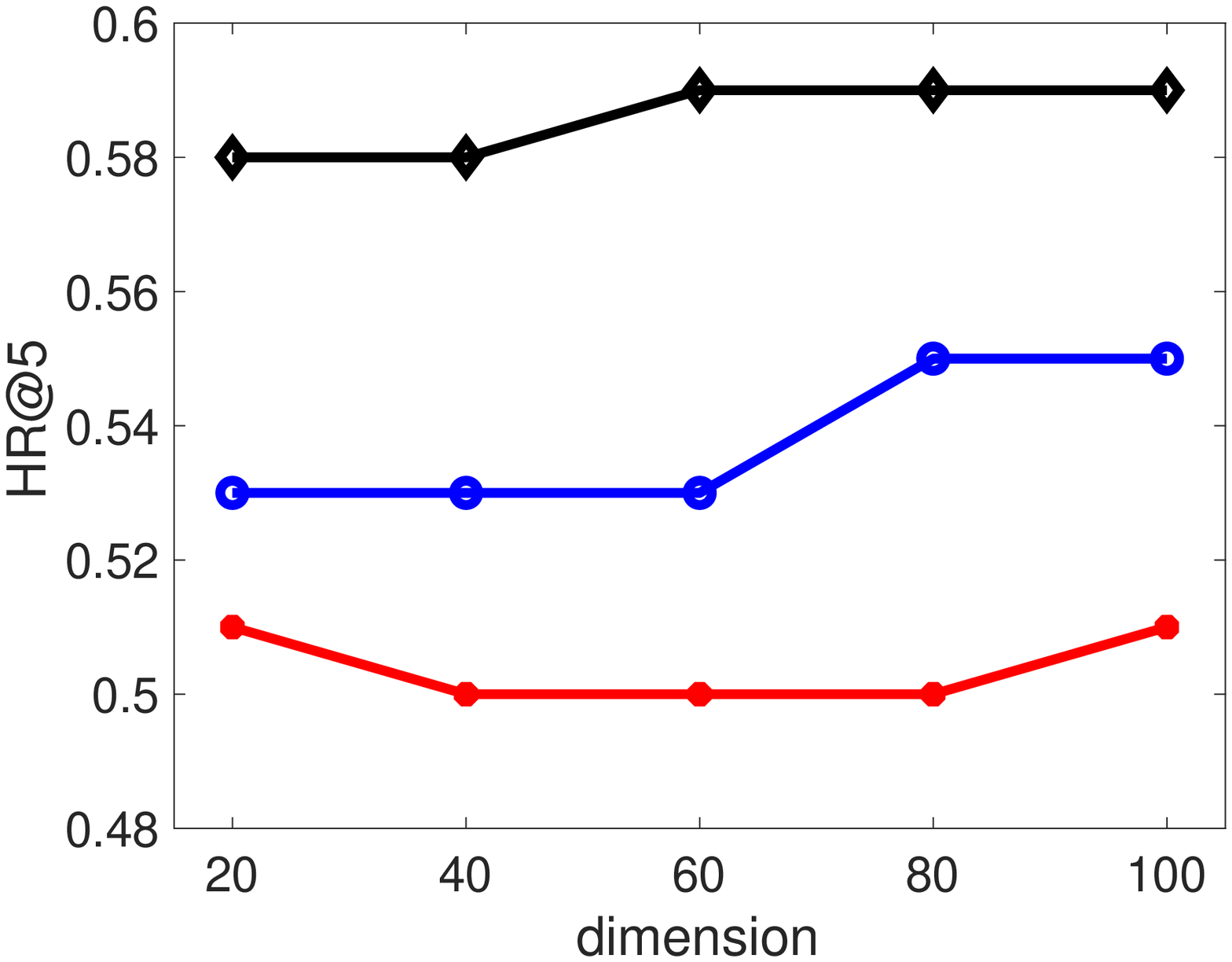} }
	\subfloat[HR@10]{\includegraphics[width=0.24\textwidth]{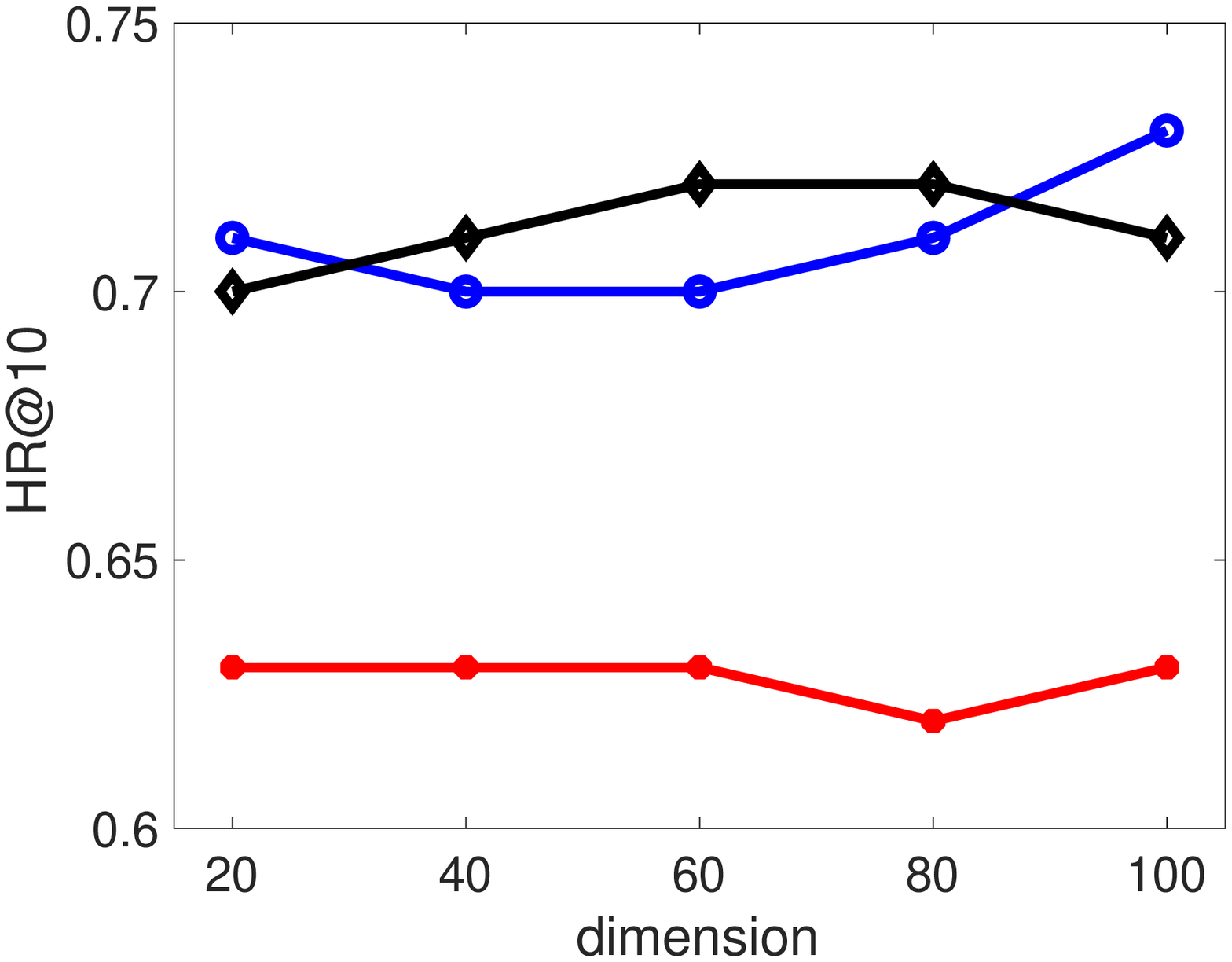} } \\
	\subfloat[NDCG@5]{\includegraphics[width=0.24\textwidth]{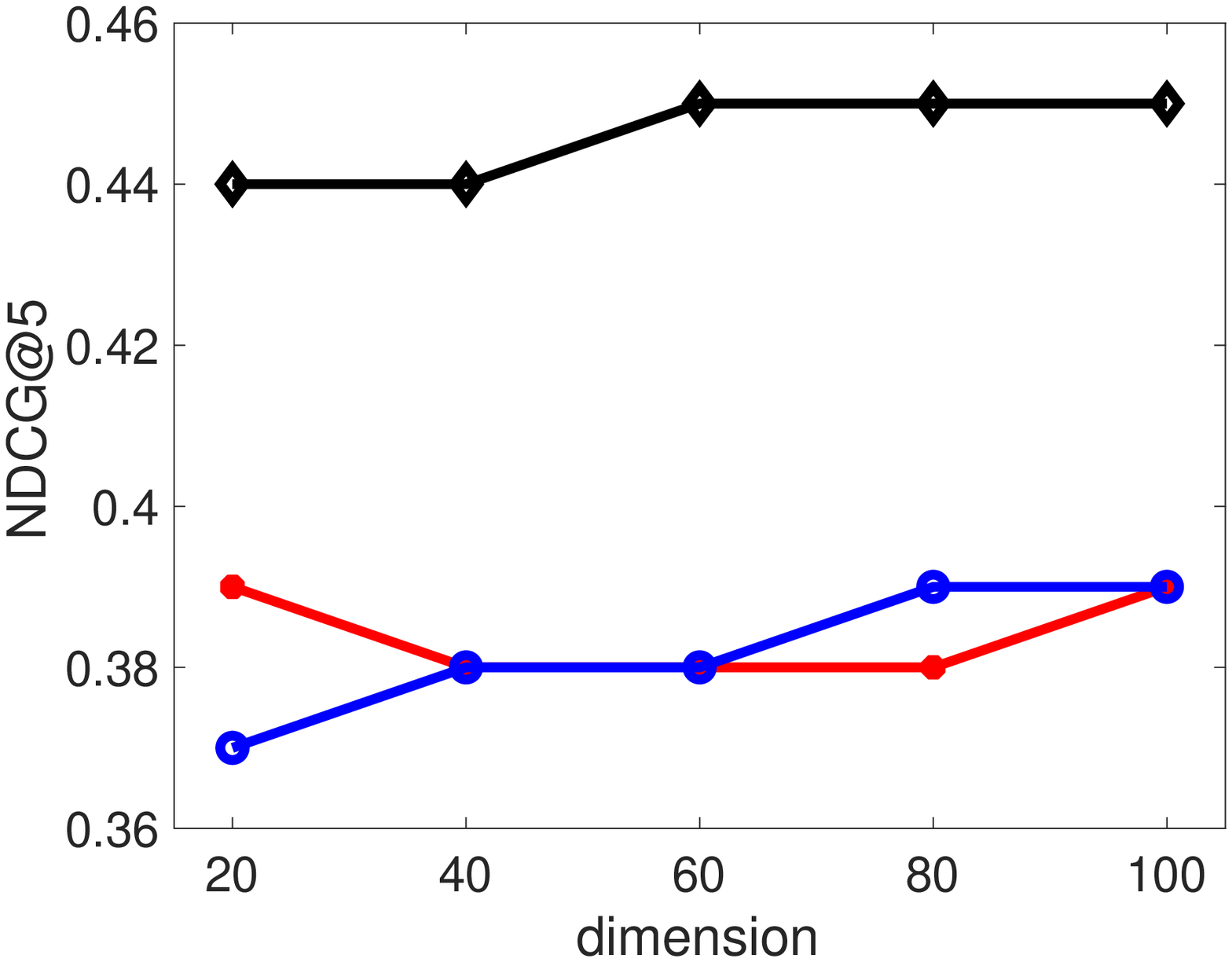} }
	\subfloat[NDCG@10]{\includegraphics[width=0.24\textwidth]{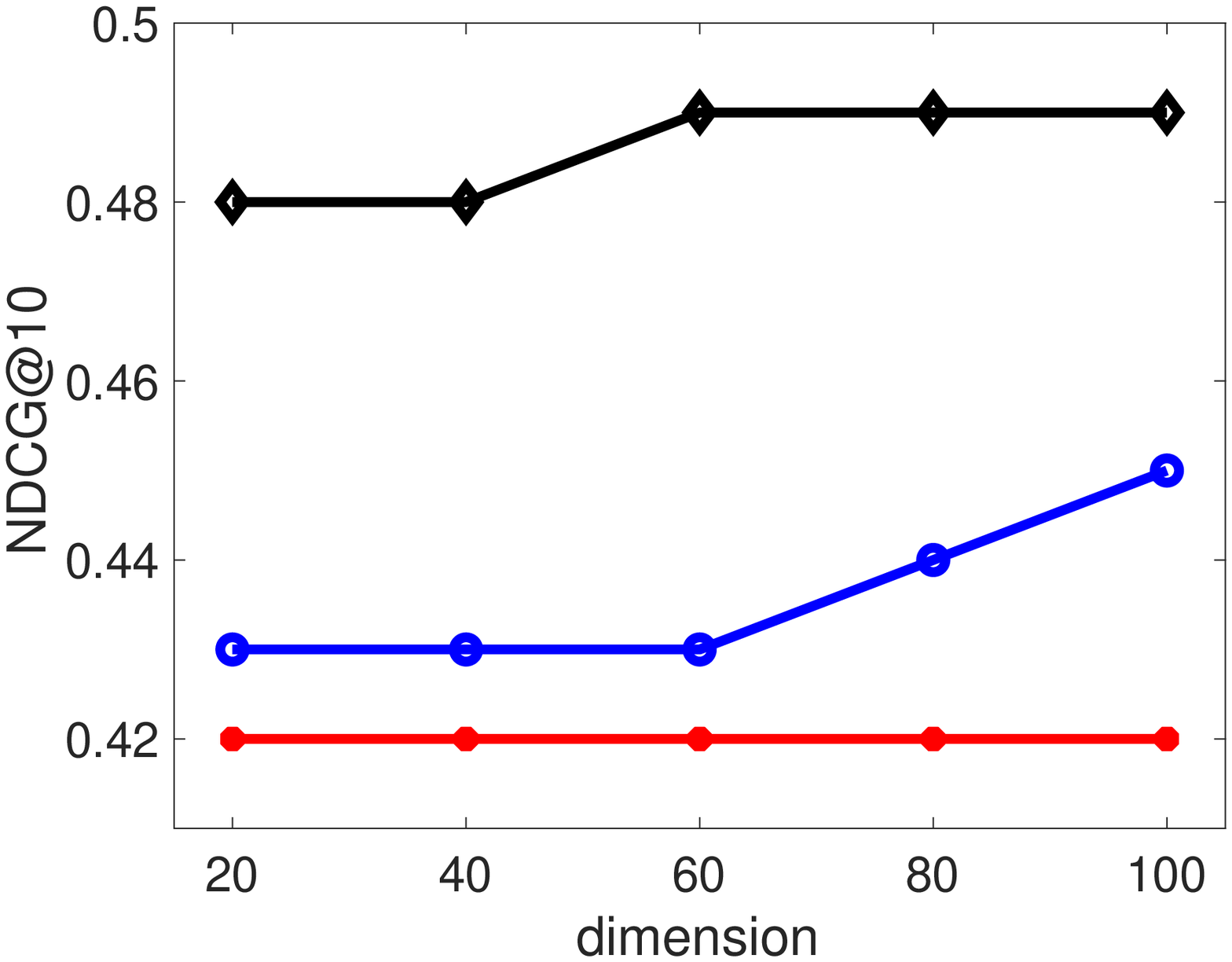} }\\
	\caption{Ranking performance under different number of embedding dimension.}
	\label{fig:dimension}
\end{figure}

\vspace{0.5em}
\noindent \textbf{Embedding dimension}:
To determine the effect of dimension of embedding latent vectors, we conduct a comparative experiment on the proposed \M.
We focus on discussing the number of embedding vectors of users, items and the fusion of events.
We report the effect of latent dimensionality on the ranking performance over the three datasets.
Fig. \ref{fig:dimension} demonstrates the validation performance under different metrics including HR@5, HR@10, NDCG@5 and NDCG@10, w.r.t. different number of embedding dimension.

Comparing the performance results under different metrics, we find that our model typically benefits from lager number of dimensionality in most cases, and achieve satisfactory performance with $d\geq60$.
Furthermore, Fig. \ref{fig:dimension} shows that \M~mostly achieves the best performance on MovieLens100K dataset, and better performance on Movie\&TV dataset, compared with the performance on Electronics dataset.
This finding reports that our proposed \M~typically obtains better on dense dataset with moderate size.
Comparing the upper and lower sides in Fig.\ref{fig:dimension}, we can also observe that the performance of \M~in terms of HR are more superior than those of NDCG.

\begin{figure}[!t]
	\centering
	\subfloat{\includegraphics[width=0.25\textwidth]{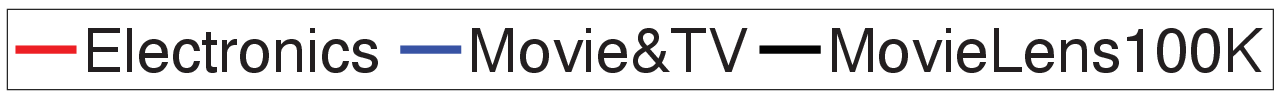} }\\
	\setcounter{subfigure}{0}
	\subfloat[HR@5]{\includegraphics[width=0.245\textwidth]{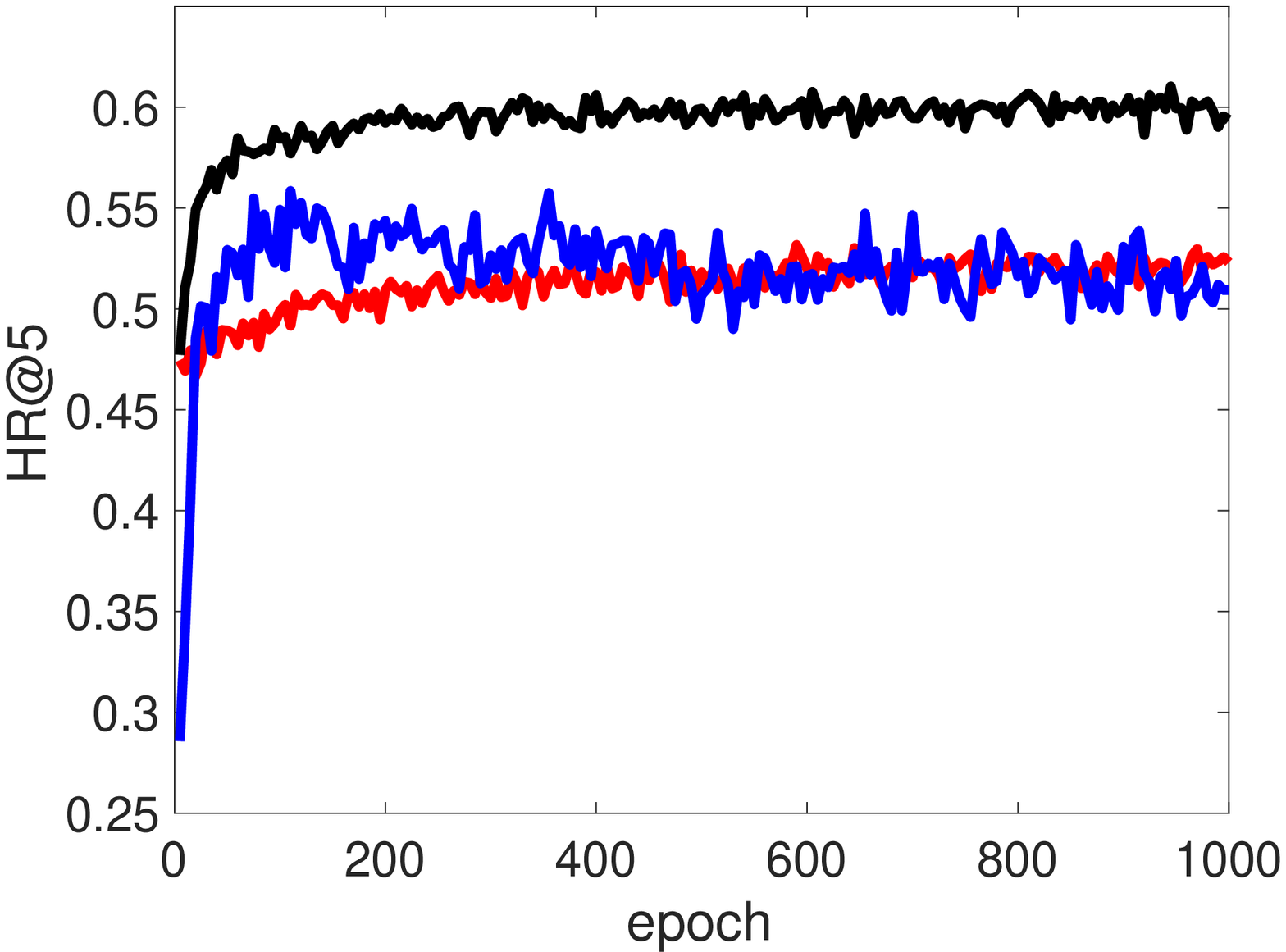} }
	\subfloat[HR@10]{\includegraphics[width=0.245\textwidth]{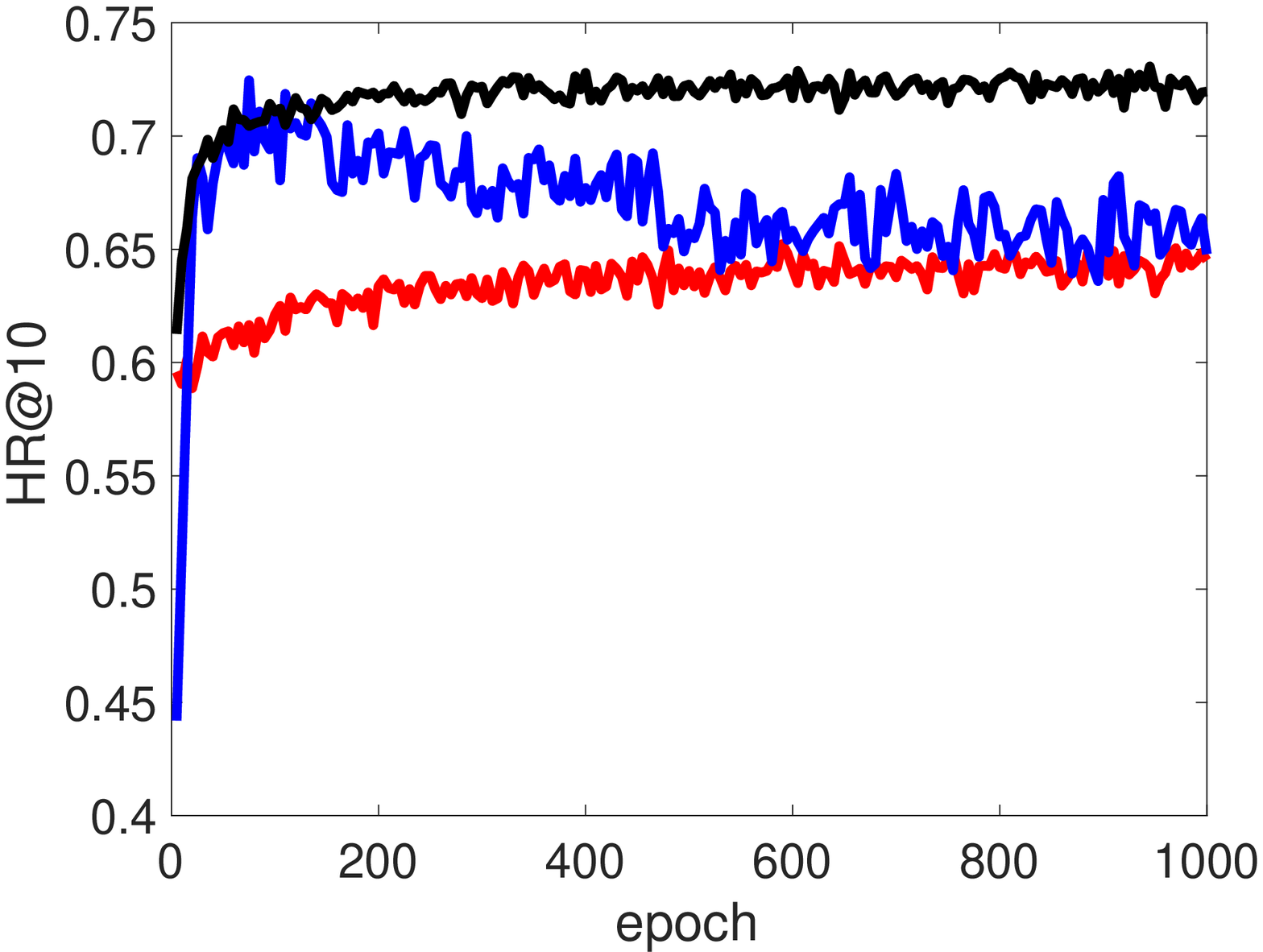} } \\
	\subfloat[NDCG@5]{\includegraphics[width=0.245\textwidth]{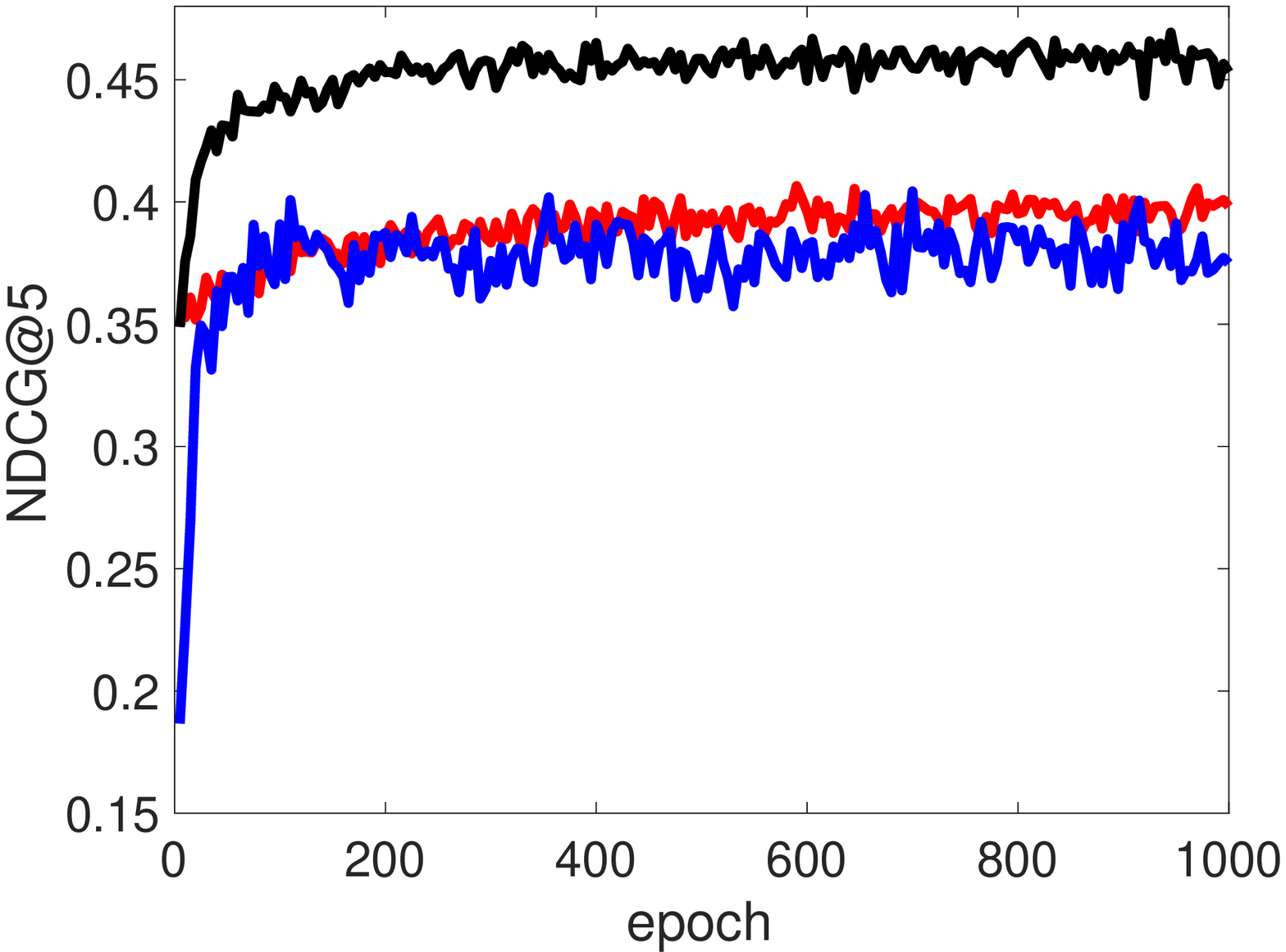} }
	\subfloat[NDCG@10]{\includegraphics[width=0.245\textwidth]{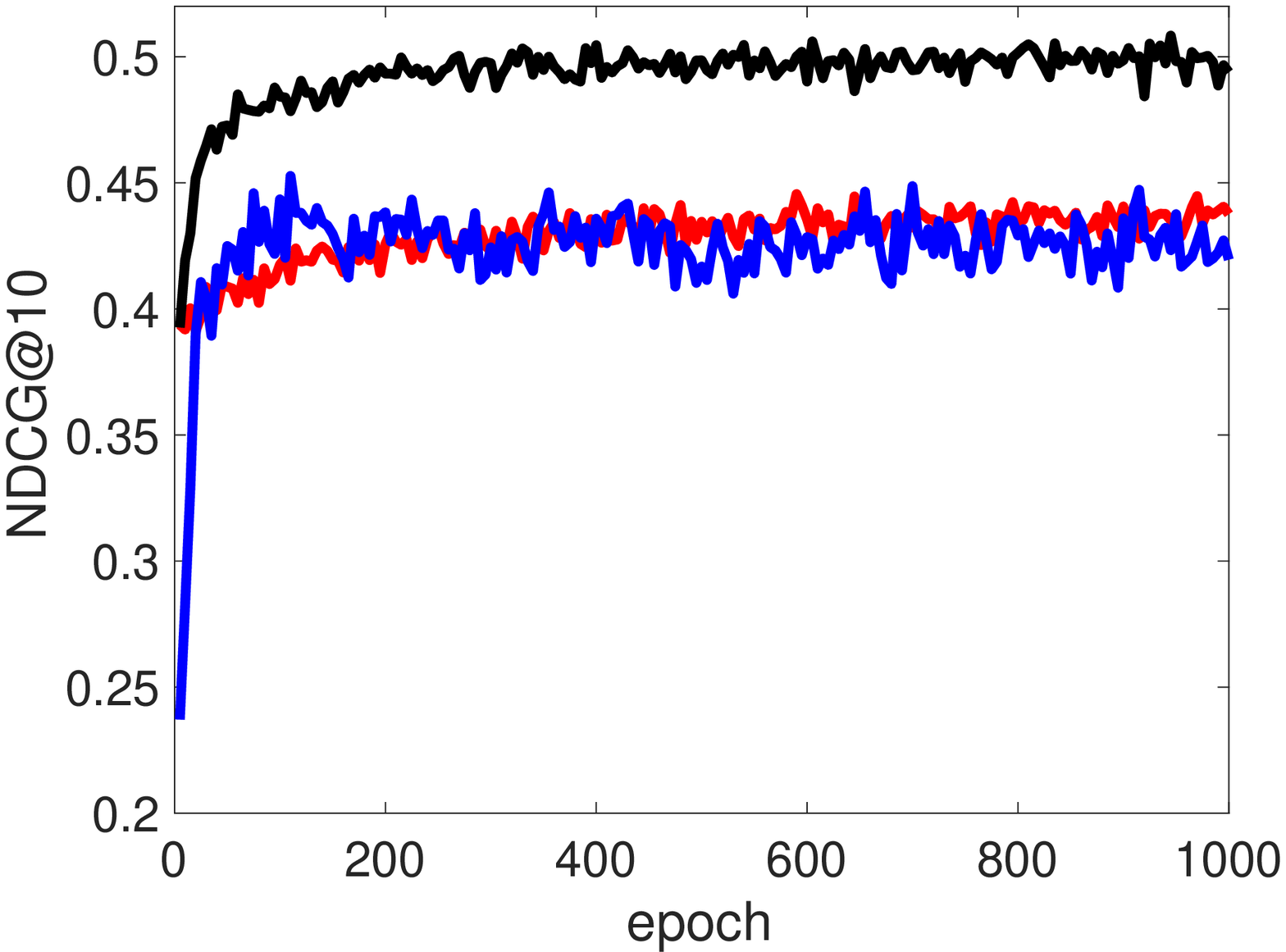} }\\
	\caption{Ranking performance across different numbers of training epochs.}
	\label{fig:epoch}
\end{figure}

\vspace{0.5em}
\noindent \textbf{Number of epoch}:
In order to investigate the effect of training epoch on the ranking performance of \M,
we conduct experiments on the training of \M~with different numbers of training epoches, which is selected in range of $5$ to $1,000$.
Fig. \ref{fig:epoch} shows the ranking performance over the three datasets across different numbers of training epochs.
In Fig. \ref{fig:epoch}, we can see that as the number of training epoch increases, the recommendation performance of \M~over all datasets is gradually improved and then achieves convergence.
The oscillations of performance results is due to the randomness of sampling in model training.

For MovieLens100K dataset, the ranking performance converges to $0.6$ over HR@5, $0.73$ over HR@10, $0.45$ over NDCG@5 and $0.5$ over NDCG@10.
For Electronics dataset, the ranking performance converges to $0.52$ over HR@5, $0.65$ over HR@10, $0.4$ over NDCG@5 and $0.43$ over NDCG@10.
For Movie\&TV dataset, the ranking performance achieves best result of $0.55$, $0.73$, $0.4$ and $0.45$ over HR@5, HR@10, NDCG@5 and NDCG@10 respectively, and then degrades after reaching the peak.
For all datasets, our \M~model obtains satisfactory performance with training epoch larger than 100.

\begin{figure}[!t]
	\centering
	\subfloat{\includegraphics[width=0.25\textwidth]{./image/legend-dimension} }\\
	\setcounter{subfigure}{0}
	\subfloat[\text{HR@$k$}]{\includegraphics[width=0.24\textwidth]{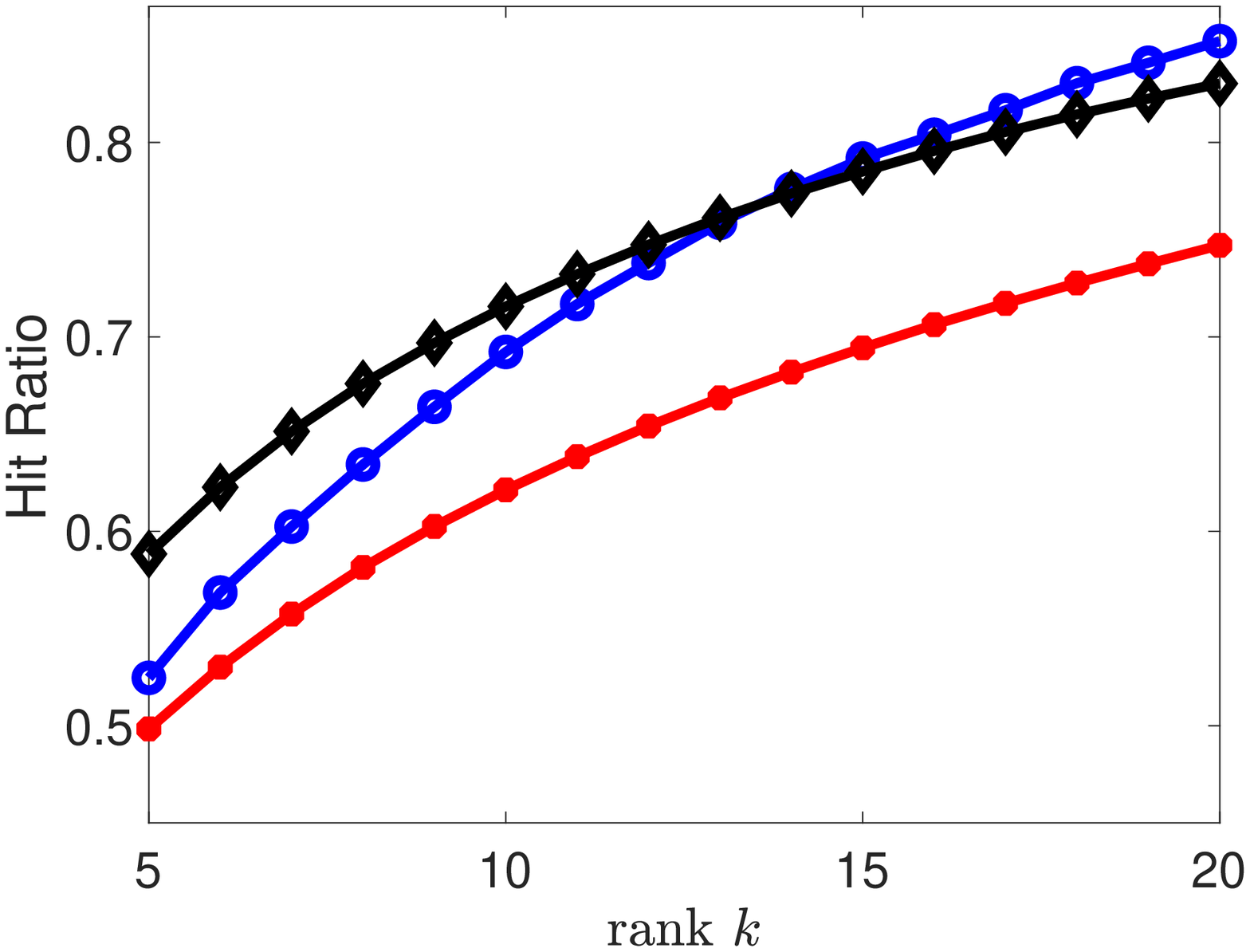} }
	\subfloat[\text{NDCG@$k$}]{\includegraphics[width=0.24\textwidth]{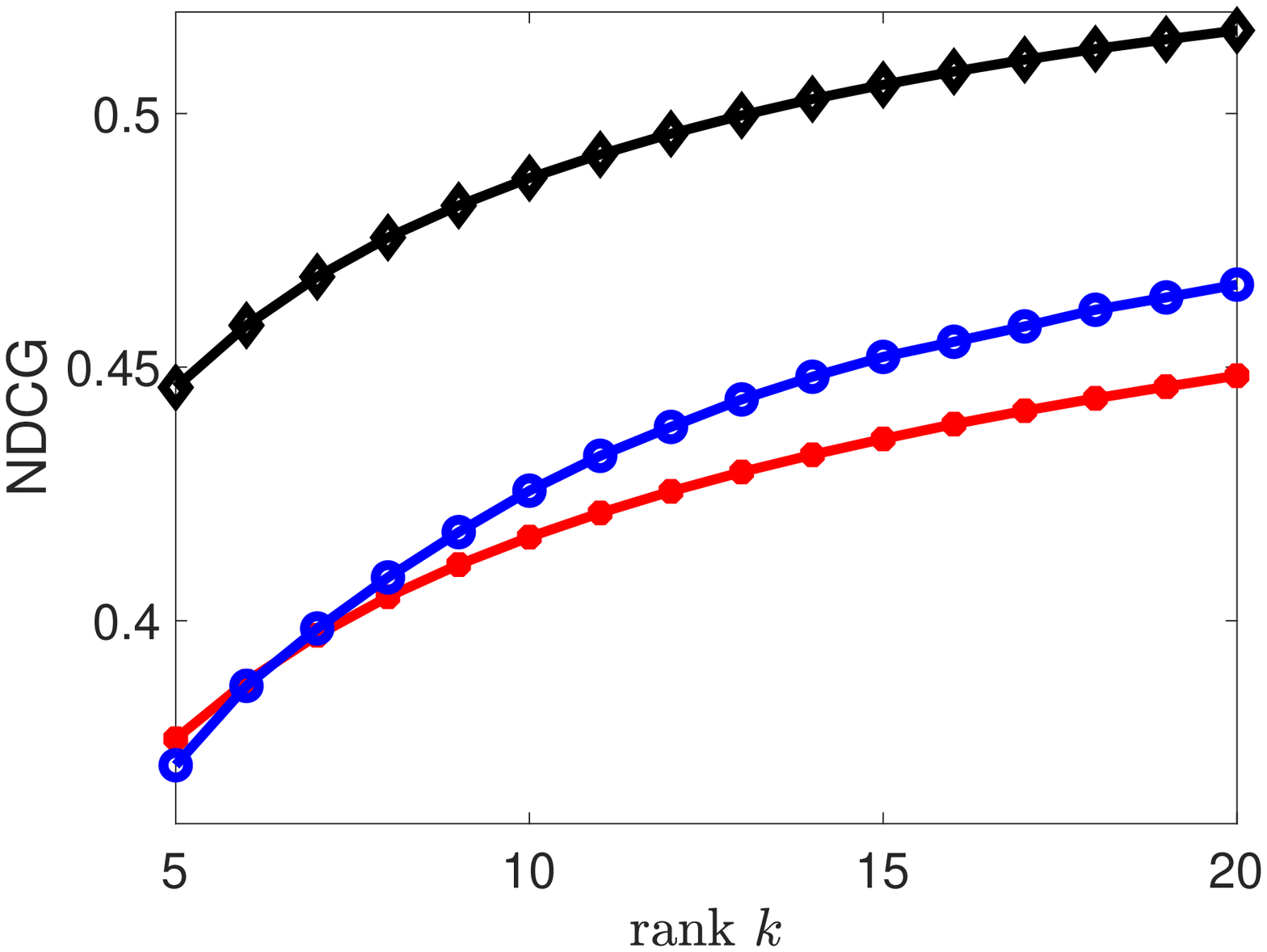} }\\
	\caption{Performance under HR and NDCG at different rank $k$, where $k$ ranges from 5 to 20. The left and right plots demonstrate the ranking performance in term of HR@$k$ and NDCG@$k$, respectively.}
	\label{fig:metric}
\end{figure}

\vspace{0.5em}
\noindent \textbf{Metrics at different rank}:
We consider to investigate the recommendation performance of \M~under HR and NDCG at different rank $k$, where $k$ ranges from 5 to 20.
Fig. \ref{fig:metric} manifests the evaluation performance of proposed \M~ under metrics HR and NDCG.
It is obvious that, with the size of rank $k$ list ranging from $5$ to $20$, the overall performance of recommendation increases, which is consistent with the conclusion in the literature of RS.
Comparing the results in Fig. \ref{fig:metric}(a) and (b), the performance of \M~in terms of HR are more superior than those in terms of NDCG.
In Fig. \ref{fig:metric}(b), under NDCG metric with different rank, the recommendation performance over MovieLens100K dataset is better than thoes over other two datasets, which also indicates that \M~typically obtains better on dense dataset with moderate size.
In Fig. \ref{fig:metric}(a), we can observe that under HR metric with different rank, the performance of \M~over Electronics dataset is worse than those of other two datasets.


\section{Conclusion and Future work} \label{sec_conclusion}
In this paper, we aim to leverage the temporal information and self-attention mechanism to capture deeper representation of user's preference for recommender system.
A novel recommendation model named \M~is proposed, where relative time patterns is employed to provide more context information, while self-attention is adopted to capture informative temporal context and suppress irrelevant patterns in historical interactions, leading to better representation of the temporal information for candidate item recommendation.
We conduct extensive experiments on three benchmark datasets to evaluate the effectiveness of \M~on modeling temporal patterns.
The experimental results demonstrate that our \M~outperforms the state-of-the-art recommendation approaches.

The incorporation of temporal information and self-attention mechanism into reasoning-based recommendation model is quite simple and straightforward but obtains significant improvement over original NCR.
This simple incorporation provides possibilities to consider extra context to enhance the representation ability of recommendation model, which will be considered as a future work.
Furthermore, the relationship between users and those between items motivate us to consider employing graph neural networks (GNN) in reasoning-based recommendation, where GNN is taken into consideration to model the complex structures of users and items.

\section*{Acknowledgment}
The authors would like to thank the anonymous reviewers for their thoughtful suggestions that would help to improve this paper substantially.



\begin{thebibliography}{10}

\bibitem{bobadilla2013recommender}
Jes{\'u}s Bobadilla, Fernando Ortega, Antonio Hernando, and Abraham
  Guti{\'e}rrez.
\newblock Recommender systems survey.
\newblock {\em Know-based Syst.}, 46:109--132, 2013.

\bibitem{campos2014time}
Pedro~G Campos, Fernando D{\'\i}ez, and Iv{\'a}n Cantador.
\newblock Time-aware recommender systems: a comprehensive survey and analysis
  of existing evaluation protocols.
\newblock {\em User Modeling and User-Adapted Interaction}, 24(1):67--119,
  2014.

\bibitem{chen2018neural}
Chong Chen, Min Zhang, Yiqun Liu, and Shaoping Ma.
\newblock Neural attentional rating regression with review-level explanations.
\newblock In {\em Proc. Conf. World Wide Web}, pages 1583--1592, 2018.

\bibitem{chen2021neural}
Hanxiong Chen, Shaoyun Shi, Yunqi Li, and Yongfeng Zhang.
\newblock Neural collaborative reasoning.
\newblock In {\em Proceedings of the Web Conference 2021}, pages 1516--1527,
  2021.

\bibitem{chen2017attentive}
Jingyuan Chen, Hanwang Zhang, Xiangnan He, Liqiang Nie, Wei Liu, and Tat-Seng
  Chua.
\newblock Attentive collaborative filtering: Multimedia recommendation with
  item-and component-level attention.
\newblock In {\em Proc. Int. ACM SIGIR Conf. Res. Dev. Inf. Retr.}, pages
  335--344, 2017.

\bibitem{dacrema2021troubling}
Maurizio~Ferrari Dacrema, Simone Boglio, Paolo Cremonesi, and Dietmar Jannach.
\newblock A troubling analysis of reproducibility and progress in recommender
  systems research.
\newblock {\em ACM Trans. Inf. Syst.}, 39(2):1--49, 2021.

\bibitem{dacrema2019we}
Maurizio~Ferrari Dacrema, Paolo Cremonesi, and Dietmar Jannach.
\newblock Are we really making much progress? a worrying analysis of recent
  neural recommendation approaches.
\newblock In {\em Proc. ACM Int. Conf. Recomm. Syst.}, pages 101--109, 2019.

\bibitem{fan2021continuous}
Ziwei Fan, Zhiwei Liu, Jiawei Zhang, Yun Xiong, Lei Zheng, and Philip~S Yu.
\newblock Continuous-time sequential recommendation with temporal graph
  collaborative transformer.
\newblock In {\em Proc. ACM Int. Conf. Inf. Knowl. Manag.}, pages 433--442,
  2021.

\bibitem{ferrari2020critically}
Maurizio Ferrari~Dacrema, Federico Parroni, Paolo Cremonesi, and Dietmar
  Jannach.
\newblock Critically examining the claimed value of convolutions over user-item
  embedding maps for recommender systems.
\newblock In {\em Proc. ACM Int. Conf. Inf. Knowl. Manag.}, pages 355--363,
  2020.

\bibitem{gantz2012digital}
John Gantz and David Reinsel.
\newblock The digital universe in 2020: Big data, bigger digital shadows, and
  biggest growth in the far east.
\newblock {\em Proc. IDC iView: IDC Analyze the future}, 2007(2012):1--16,
  2012.

\bibitem{harper2015movielens}
F~Maxwell Harper and Joseph~A Konstan.
\newblock The movielens datasets: History and context.
\newblock {\em ACM Trans. Interact. Intell. Syst.}, 5(4):1--19, 2015.

\bibitem{he2017neural}
Xiangnan He, Lizi Liao, Hanwang Zhang, Liqiang Nie, Xia Hu, and Tat-Seng Chua.
\newblock Neural collaborative filtering.
\newblock In {\em Proc. Int. Conf. World Wide Web}, pages 173--182, 2017.

\bibitem{hidasi2015session}
Bal{\'a}zs Hidasi, Alexandros Karatzoglou, Linas Baltrunas, and Domonkos Tikk.
\newblock Session-based recommendations with recurrent neural networks.
\newblock {\em arXiv preprint arXiv:1511.06939}, 2015.

\bibitem{ji2020sequential}
Wendi Ji, Keqiang Wang, Xiaoling Wang, Tingwei Chen, and Alexandra Cristea.
\newblock Sequential recommender via time-aware attentive memory network.
\newblock In {\em Proc. ACM Int. Conf. Inf. Knowl. Manag.}, pages 565--574,
  2020.

\bibitem{kang2018self}
Wang-Cheng Kang and Julian McAuley.
\newblock Self-attentive sequential recommendation.
\newblock In {\em Proc. IEEE Int. Conf. Data Mining}, pages 197--206, 2018.

\bibitem{koenigstein2011yahoo}
Noam Koenigstein, Gideon Dror, and Yehuda Koren.
\newblock Yahoo! music recommendations: modeling music ratings with temporal
  dynamics and item taxonomy.
\newblock In {\em Proc. Conf. Recomm. Syst.}, pages 165--172, 2011.

\bibitem{koren2008factorization}
Yehuda Koren.
\newblock Factorization meets the neighborhood: a multifaceted collaborative
  filtering model.
\newblock In {\em Proc. Int. Conf. Knowl. Discov. Data Mining}, pages 426--434,
  2008.

\bibitem{koren2009collaborative}
Yehuda Koren.
\newblock Collaborative filtering with temporal dynamics.
\newblock In {\em roc. ACM SIGKDD Int. Conf. Knowl. Disc. Data Mining}, pages
  447--456, 2009.

\bibitem{koren2009matrix}
Yehuda Koren, Robert Bell, and Chris Volinsky.
\newblock Matrix factorization techniques for recommender systems.
\newblock {\em Computer}, 42(8):30--37, 2009.

\bibitem{kumar2019predicting}
Srijan Kumar, Xikun Zhang, and Jure Leskovec.
\newblock Predicting dynamic embedding trajectory in temporal interaction
  networks.
\newblock In {\em Proc. ACM SIGKDD Int. Conf. Knowl. Disc. Data Mining}, pages
  1269--1278, 2019.

\bibitem{li2020time}
Jiacheng Li, Yujie Wang, and Julian McAuley.
\newblock Time interval aware self-attention for sequential recommendation.
\newblock In {\em Proc. Int. Conf. Web Search Data Dining}, pages 322--330,
  2020.

\bibitem{li2014modeling}
Lei Li, Li~Zheng, Fan Yang, and Tao Li.
\newblock Modeling and broadening temporal user interest in personalized news
  recommendation.
\newblock {\em Expert Systems with Applications}, 41(7):3168--3177, 2014.

\bibitem{liu2010personalized}
Jiahui Liu, Peter Dolan, and Elin~R{\o}nby Pedersen.
\newblock Personalized news recommendation based on click behavior.
\newblock In {\em Proc. Int. Conf. Intell. User Interfaces}, pages 31--40,
  2010.

\bibitem{liu2018stamp}
Qiao Liu, Yifu Zeng, Refuoe Mokhosi, and Haibin Zhang.
\newblock Stamp: short-term attention/memory priority model for session-based
  recommendation.
\newblock In {\em Proc. ACM SIGKDD Int. Conf. Knowl. Disc. Data Mining}, pages
  1831--1839, 2018.

\bibitem{lu2015recommender}
Jie Lu, Dianshuang Wu, Mingsong Mao, Wei Wang, and Guangquan Zhang.
\newblock Recommender system application developments: a survey.
\newblock {\em Decis. Support Syst.}, 74:12--32, 2015.

\bibitem{lu2012recommender}
Linyuan L{\"u}, Mat{\'u}{\v{s}} Medo, Chi~Ho Yeung, Yi-Cheng Zhang, Zi-Ke
  Zhang, and Tao Zhou.
\newblock Recommender systems.
\newblock {\em Physics reports}, 519(1):1--49, 2012.

\bibitem{luong2015effective}
Minh-Thang Luong, Hieu Pham, and Christopher~D Manning.
\newblock Effective approaches to attention-based neural machine translation.
\newblock {\em arXiv preprint arXiv:1508.04025}, 2015.

\bibitem{mcauley2015image}
Julian McAuley, Christopher Targett, Qinfeng Shi, and Anton Van Den~Hengel.
\newblock Image-based recommendations on styles and substitutes.
\newblock In {\em Proc. Int. ACM SIGIR Con. Res. Dev. Info. Retr.}, pages
  43--52, 2015.

\bibitem{pei2017interacting}
Wenjie Pei, Jie Yang, Zhu Sun, Jie Zhang, Alessandro Bozzon, and David~MJ Tax.
\newblock Interacting attention-gated recurrent networks for recommendation.
\newblock In {\em Proc. ACM Conf. Inf. Knowl. Manag.}, pages 1459--1468, 2017.

\bibitem{rendle2012bpr}
Steffen Rendle, Christoph Freudenthaler, Zeno Gantner, and Lars Schmidt-Thieme.
\newblock Bpr: Bayesian personalized ranking from implicit feedback.
\newblock {\em arXiv preprint arXiv:1205.2618}, 2012.

\bibitem{resnick1994grouplens}
Paul Resnick, Neophytos Iacovou, Mitesh Suchak, Peter Bergstrom, and John
  Riedl.
\newblock Grouplens: An open architecture for collaborative filtering of
  netnews.
\newblock In {\em Proc. ACM Int. Conf. Computer supported cooperative work},
  pages 175--186, 1994.

\bibitem{sedhain2015autorec}
Suvash Sedhain, Aditya~Krishna Menon, Scott Sanner, and Lexing Xie.
\newblock Auto{R}ec: Autoencoders meet collaborative filtering.
\newblock In {\em Proc. ACM Int. Conf. World Wide Web}, pages 111--112, 2015.

\bibitem{shaw2018self}
Peter Shaw, Jakob Uszkoreit, and Ashish Vaswani.
\newblock Self-attention with relative position representations.
\newblock In {\em NAACL-HLT (2)}, 2018.

\bibitem{shi2020neural}
Shaoyun Shi, Hanxiong Chen, Weizhi Ma, Jiaxin Mao, Min Zhang, and Yongfeng
  Zhang.
\newblock Neural logic reasoning.
\newblock In {\em Proc. ACM Int. Conf. Inf. Knowl. Manag.}, pages 1365--1374,
  2020.

\bibitem{song2016multi}
Yang Song, Ali~Mamdouh Elkahky, and Xiaodong He.
\newblock Multi-rate deep learning for temporal recommendation.
\newblock In {\em Proc. Int. ACM SIGIR Conf. Res. Dev. Inf. Retr.}, pages
  909--912, 2016.

\bibitem{sun2018attentive}
Peijie Sun, Le~Wu, and Meng Wang.
\newblock Attentive recurrent social recommendation.
\newblock In {\em Int. ACM SIGIR Conf. Res. Dev. Inf. Retr.}, pages 185--194,
  2018.

\bibitem{wang2017residual}
Fei Wang, Mengqing Jiang, Chen Qian, Shuo Yang, Cheng Li, Honggang Zhang,
  Xiaogang Wang, and Xiaoou Tang.
\newblock Residual attention network for image classification.
\newblock In {\em Proc. Int. Conf. Comp. Vis. Patt. Recog.}, pages 3156--3164,
  2017.

\bibitem{wang2015collaborative}
Hao Wang, Naiyan Wang, and Dit-Yan Yeung.
\newblock Collaborative deep learning for recommender systems.
\newblock In {\em Proc. Int. Conf. Knowl. Discov. Data Mining}, pages
  1235--1244, 2015.

\bibitem{wu2017recurrent}
Chao-Yuan Wu, Amr Ahmed, Alex Beutel, Alexander~J Smola, and How Jing.
\newblock Recurrent recommender networks.
\newblock In {\em Proc. ACM Int. Conf. Web Search Data Mining}, pages 495--503,
  2017.

\bibitem{wu2020deja}
Jibang Wu, Renqin Cai, and Hongning Wang.
\newblock D{\'e}j{\`a} vu: A contextualized temporal attention mechanism for
  sequential recommendation.
\newblock In {\em Proceedings of The Web Conference}, pages 2199--2209, 2020.

\bibitem{wu2016collaborative}
Yao Wu, Christopher DuBois, Alice~X Zheng, and Martin Ester.
\newblock Collaborative denoising auto-encoders for top-n recommender systems.
\newblock In {\em Proc. Int. Conf. Web Search Data Mining}, pages 153--162,
  2016.

\bibitem{xiao2017attentional}
Jun Xiao, Hao Ye, Xiangnan He, Hanwang Zhang, Fei Wu, and Tat-Seng Chua.
\newblock Attentional factorization machines: learning the weight of feature
  interactions via attention networks.
\newblock In {\em Proc. Int. Joint Conf. Artif. Intell.}, pages 3119--3125,
  2017.

\bibitem{xiong2010temporal}
Liang Xiong, Xi~Chen, Tzu-Kuo Huang, Jeff Schneider, and Jaime~G Carbonell.
\newblock Temporal collaborative filtering with bayesian probabilistic tensor
  factorization.
\newblock In {\em Pro. Int. Conf. Data Mining}, pages 211--222, 2010.

\bibitem{xue2017deep}
Hong-Jian Xue, Xinyu Dai, Jianbing Zhang, Shujian Huang, and Jiajun Chen.
\newblock Deep matrix factorization models for recommender systems.
\newblock In {\em Int. Joint Conf Artif. Intell.}, volume~17, pages 3203--3209.
  Melbourne, Australia, 2017.

\bibitem{ye2020time}
Wenwen Ye, Shuaiqiang Wang, Xu~Chen, Xuepeng Wang, Zheng Qin, and Dawei Yin.
\newblock Time matters: Sequential recommendation with complex temporal
  information.
\newblock In {\em Proc. Int. ACM SIGIR Conf. Res. Dev. Inf. Retr.}, pages
  1459--1468, 2020.

\bibitem{ying2018sequential}
Haochao Ying, Fuzhen Zhuang, Fuzheng Zhang, Yanchi Liu, Guandong Xu, Xing Xie,
  Hui Xiong, and Jian Wu.
\newblock Sequential recommender system based on hierarchical attention
  network.
\newblock In {\em Int. Joint Conf. Artif. Intell.}, 2018.

\bibitem{yu2016dynamic}
Feng Yu, Qiang Liu, Shu Wu, Liang Wang, and Tieniu Tan.
\newblock A dynamic recurrent model for next basket recommendation.
\newblock In {\em Proc. Int. ACM SIGIR Conf. Res. Dev. Inf. Retr.}, pages
  729--732, 2016.

\bibitem{yu2016temporal}
Hsiang-Fu Yu, Nikhil Rao, and Inderjit~S Dhillon.
\newblock Temporal regularized matrix factorization for high-dimensional time
  series prediction.
\newblock {\em Adv. Neural Inf. Process. Syst.}, 29, 2016.

\bibitem{yuan2013time}
Quan Yuan, Gao Cong, Zongyang Ma, Aixin Sun, and Nadia~Magnenat Thalmann.
\newblock Time-aware point-of-interest recommendation.
\newblock In {\em Proc. Int. ACM SIGIR Conf. Rese. Dev. Inf. Retr.}, pages
  363--372, 2013.

\bibitem{zhai2006visual}
Yun Zhai and Mubarak Shah.
\newblock Visual attention detection in video sequences using spatiotemporal
  cues.
\newblock In {\em Proc. Int. Conf. Multimedia}, pages 815--824, 2006.

\bibitem{zhang2019next}
Shuai Zhang, Yi~Tay, Lina Yao, Aixin Sun, and Jake An.
\newblock Next item recommendation with self-attentive metric learning.
\newblock In {\em AAAI Conf. Artif. Intell.}, volume~9, 2019.

\bibitem{zheng2017joint}
Lei Zheng, Vahid Noroozi, and Philip~S Yu.
\newblock Joint deep modeling of users and items using reviews for
  recommendation.
\newblock In {\em Proc. Int. Conf. Knowl. Discov. Data Mining}, pages 425--434,
  2017.

\end{thebibliography}

\end{document}